\documentclass[a4paper,11pt]{article}
\usepackage{jcappub} 
\usepackage{lineno}
\usepackage{xcolor}
\usepackage{float}

\usepackage{makecell}
\usepackage{multirow}
\usepackage{tablefootnote}
\usepackage{array}
\usepackage{silence}
\arxivnumber{2408.14628} 
\title{Investigating Late-Time Dark Energy and Massive Neutrinos in Light of DESI Y1 BAO}

\author[a]{João Rebouças,}
\author[a]{Diogo H. F. de Souza,}
\author[b]{Kunhao Zhong,}
\author[c]{Vivian Miranda}
\author[a]{and Rogerio Rosenfeld}
\affiliation[a]{Instituto de Física Teórica da Universidade Estadual Paulista and ICTP South American Institute for Fundamental Research, R. Dr. Bento Teobaldo Ferraz, 271, Bloco II, Barra-Funda - Sao Paulo/SP, Brasil}
\affiliation[b]{Department of Physics and Astronomy, University of Pennsylvania, Philadelphia, PA 19104, USA}
\affiliation[c]{C.N. Yang Institute for Theoretical Physics, Stony Brook University, NY 11794, USA}

\emailAdd{joao.reboucas@unesp.br}
\emailAdd{diogo.henrique@unesp.br}

\abstract{Baryonic Acoustic Oscillation (BAO) data from the Dark Energy Spectroscopic Instrument (DESI), in combination with Cosmic Microwave Background (CMB) data and Type Ia Supernovae (SN) luminosity distances, suggests a dynamical evolution of the dark energy equation of state with a phantom phase ($w < -1$) in the past when the so-called $w_0w_a$ parametrization $w(a) = w_0 + w_a(1-a)$ is assumed. In this work, we investigate more general dark energy models that also allow a phantom equation of state. We consider three cases: an equation of state with a transition feature, a model-agnostic equation of state with constant values in chosen redshift bins, and a k-essence model. Since the dark energy equation of state is correlated with neutrino masses, we reassess constraints on the neutrino mass sum focusing on the model-agnostic equation of state. We find that the combination of DESI BAO with Planck 2018 CMB data and SN data from Pantheon, Pantheon+, or Union3 is consistent with an oscillatory dark energy equation of state, while a monotonic behavior is preferred by the DESY5 SN data. Performing model comparison techniques, we find that the $w_0w_a$ parametrization remains the simplest dark energy model that can provide a better fit to DESI BAO, CMB, and all SN datasets than $\Lambda$CDM. Constraints on the neutrino mass sum assuming dynamical dark energy are relaxed compared to $\Lambda$CDM and we show that these constraints are tighter in the model-agnostic case relative to $w_0w_a$ model by $70\%-90\%$.}

\begin{document}
\maketitle
\flushbottom

\section{Introduction}
\label{sec:intro}

The nature of dark energy persists as one of the greatest mysteries in current physics. Ever since the discovery of accelerated expansion in 1998 \cite{SupernovaSearchTeam:1998fmf, SupernovaCosmologyProject:1998vns}, the cosmological constant $\Lambda$ has become the accepted model of dark energy, due to its simplicity and ability to fit various independent observables: Type Ia supernovae luminosity distances \cite{snls, jla, pantheon, pantheonplus, union3, desy5, desy5_cosmo}, distance measurements from Baryonic Acoustic Oscillations (BAOs) \cite{sdss_bao, 2dfgs_bao, sdss_2fdgs_bao, sdss_boss_bao, sdss_dr10_bao, sdss_dr12_bao, sdss_dr7_bao, sdss_dr9_bao, sdss_lya_bao, sdss_lya2_bao, 6dfgs2_bao, 6dfgs3_bao, sdss4_bao, wigglez_bao, wigglez2_bao, wigglez3_bao, sdss_lya3_bao, sdss4_qso_bao, desi_2024_bao, bao_dark_energy}, Cosmic Microwave Background (CMB) temperature and polarization anisotropies \cite{cobe_cmb, wmap_cmb, firas_cmb, bicep_cmb, planck_2015_cmb, planck_2018_cmb, act_cmb, spt_cmb}, two-point correlation functions of galaxy shapes and positions \cite{apm_shear, hsc-shear, cfhtlens_shear, kids-1000-shear, des_y1_3x2, des_y3_3x2}, among other probes. The success of the $\Lambda$CDM model, which assumes a cosmological constant as dark energy, has ushered in an era of precision cosmology, in which increasingly ambitious surveys can track the expansion of the Universe and the growth of large-scale structure with great precision. However, the $\Lambda$CDM model faces several challenges. From a theoretical standpoint, its connection with the Standard Model of Particle Physics (SM) is not well understood: the dark energy density is orders of magnitude lower than the vacuum energy density of the SM fields \cite{RevModPhys.61.1, vacuum_energy}. Furthermore, in recent years, different datasets have disagreed in their measured values of the $\Lambda$CDM cosmological parameters. The most pressing tension is in the $H_0$ parameter, the current expansion rate of the universe, between local distance measurements \cite{sh0es_hubble} and those inferred by CMB measurements (see \cite{cosmology_intertwined} for a recent review). Less statistically significant tensions, such as the $S_8$ parameter tension between CMB and cosmic shear surveys \cite{reanalysis_stage_iii_shear, kids-1000-shear}, aggravate the $\Lambda$CDM crisis. In this context, the last decade witnessed an explosion of alternative cosmological models trying to alleviate tensions and provide insights about the fundamental nature of dark energy \cite{h0_olympics, to_h0_or_not, hubble_hunters, hubble_tension_spt_sh0es, cde_hts, ede_hts, omni_hts, odyssey_hts, kessence_hts, axion_ede_hts, chameleon_hts, dissecting_hts, oscillations_hts, dataset_tensions_hts, microphysics_ede, s8_tension_nonlinear_solution, ede_ups_downs, sigma8_drag, growth_modification_s8, flamingo_s8, tensions_early_late, ht_bimetric_gravity}.

A recent hint on deviations from the $\Lambda$CDM model arising from new type Ia supernovae analyses from both Dark Energy Survey 5-year dataset (DESY5) \cite{desy5_cosmo} and Union3 \cite{union3_sn} became more prominent with the new BAO distance measurements from DESI  \cite{desi_2024_bao}. The dataset combination of DESI BAO with Planck 2018 CMB anisotropies \cite{planck_2018_cmb} and Type Ia supernovae distances from either Pantheon+ \cite{pantheonplus}, Union3 \cite{union3}, or DESY5 \cite{desy5} significantly prefers a dynamical dark energy behavior, with confidence ranging from $2.5\sigma-3.9\sigma$ depending on which supernovae catalog is being used. DESI analysis relied on the widely used $w_0w_a$ parametrization of dark energy: a phenomenological model where the equation of state evolves linearly with scale factor as $w(a) = w_0 + w_a(1-a)$. The preferred dark energy behavior in the DESI analysis has a phantom behavior ($w < -1$) \cite{phantomde} for redshifts $z > 0.5$ and a thawing behavior with $w > -1$ nowadays. The main driver of this evidence is still under debate \cite{desi_lowz_sn, desi_confirm_lcdm, desy5_omegam}. While \cite{desi_2024_bao} points that the Luminous Red Galaxy samples between redshifts $0.6 < z < 0.8$ drive this preference \cite{desi_lrg1_lrg2}, reference \cite{scalar_fields_desi} has shown that under the $\Lambda$CDM model there is a mild tension in the $\Omega_m$ parameter between DESI BAO and the Pantheon+, Union3, and DESY5 supernovae catalogs, which is alleviated by thawing dynamical dark energy. Parameter constraints using the Pantheon supernovae catalog, which agrees with DESI on $\Omega_m$, are consistent with $\Lambda$CDM. However, the Pantheon+ analysis improves over the original Pantheon on several aspects, such as an updated light curve model, which takes into account galactic dust and an improved model for intrinsic brightness evolution \cite{dust_sn}, among others. It has also been noted that the preference for an evolving dark energy equation of state depends on priors for $w_0$ and $w_a$ \cite{liddle_interpreting_desi, desi_prior_dependence, desi_self_consistency}.

These intriguing clues for dynamical dark energy resulted in a flurry of studies considering
more complex behaviors for the dark energy equation of state than the monotonic $w_0w_a$ parametrization. Some examples are scalar field quintessence models \cite{scalar_fields_desi,desi_quintessence_interpretation,assessing_constraints_desi,desi_clues, thawing_quint}, 
higher order terms in the Taylor expansion of the equation of state of dark energy around the present epoch \cite{desi_consistent_theories}, different 2-parameter descriptions of the equation of state \cite{desi_robust} and other generalizations \cite{desi_deviation_from_lcdm}, modified gravity \cite{desi_modified_grav,desi_mg_foft}, Gaussian Processes \cite{desi_gaussian_process, Jiang:2024xnu, desi_quintom}, interacting dark energy models \cite{desi_ide}, non-minimally coupled gravity \cite{desi_nonminimal_gravity}, emergent dark energy model \cite{desi_pede} and a modified recombination history \cite{desi_modified_recombination}, sterile neutrinos and modified gravity \cite{desi_physics}, a negative cosmological constant \cite{desi_negative_cc}, early dark energy \cite{desi_ede}, among others \cite{ratra_desi}.
 
A preference for dynamical dark energy over a cosmological constant also has implications for other model components. Cosmological probes have been putting increasingly stringent constraints on the neutrino mass sum \cite{nu_detecting_cdm, nu_massive_lss, nu_model_clustering, nu_massive_phenomenology, nu_mass_cosmology, nu_towards_detection, nu_power_spectra, nu_species_number, nu_weighing, model_marginalized_nu_mass}, culminating with the recent constraints from the dataset combination of Planck 2018 CMB data and DESI BAO, assuming $\Lambda$CDM: $\sum m_\nu \leq 0.072~\mathrm{eV}$ at $95\%$ confidence level (CL) \cite{desi_2024_bao}. These results are close to the lower bounds inferred from neutrino oscillation experiments, either for the normal-ordered hierarchy ($\sum m_\nu \geq 0.057~\mathrm{eV}$) or for an inverted-ordered hierarchy ($\sum m_\nu \geq 0.10~\mathrm{eV}$) \cite{nu_oscillations_review, nu_oscillations_2020, nu_oscillations_update}. One could argue that the inverted hierarchy has been ruled out in the context of the $\Lambda$CDM model.
In addition, as large-scale structure surveys obtain more precise measurements, a tension between cosmological and laboratory neutrino mass constraints could arise \cite{desi_weighing_neutrinos, neutrinos_after_desi, desi_neutrino_refractive, mnu_edge, negative_neutrino_mass, negative_nu_mass_mirage, desi_planck_pr4, desi_y_distortions_cnub, desi_unveiling_mnu}. Therefore, studying the bounds on the sum of neutrino masses in other cosmological models and datasets and their robustness under model assumptions and analysis choices is essential \cite{implications_de_massive_nu, desi_updating_mnu}. Previous studies have pointed to a preference for negative cosmological neutrino masses, which tighten the neutrino mass constraints and could be a hint for new physics \cite{negative_neutrino_mass, negative_nu_mass_mirage, no_nus_good_news}.
Several works on neutrino masses, taking into account the DESI data, appeared recently.
For example, 
\cite{mnu_edge} and \cite{negative_neutrino_mass} study the robustness of the DESI results concerning neutrino masses;
neutrino mass constraints under $\Lambda$CDM, $w$CDM and $w_0w_a$CDM and with a non-phantom $w_0w_a$CDM with the different hierarchy choices have been addressed in \cite{desi_weighing_neutrinos} and \cite{ neutrinos_after_desi},
and modifications due to new interactions between neutrinos and dark matter were considered in \cite{desi_neutrino_refractive}.

This work investigates general dark energy models using DESI BAO data alongside Type Ia supernovae distance catalogs and CMB anisotropy and lensing data. We test three models: a phenomenological model with a transition of $w(z)$ between two values, a model-agnostic parametrization in which $w(z)$ is a piecewise-constant function over given redshift bins, and a scalar field model. Quintessence models, where dark energy is related to a scalar field minimally coupled to gravity with a canonical kinetic term, are restricted to have an equation of state $w \geq -1$. To reproduce a phantom behavior, we focus on the monodromic k-essence (MKE) model introduced in \cite{mke_model}, in which the dark energy equation of state can be oscillatory. After assessing the model parameter constraints and the $w(z)$ dynamical behavior preferred by different combinations of datasets, we compare the dark energy models by how they can improve the goodness-of-fit to each dataset, penalizing models with an increased number of parameters. Furthermore, we reassess the constraints on the neutrino mass sum considering the model-agnostic parametrization of dark energy as compared to the usual $w_0w_a$CDM model, 
obtaining constraints for $\sum m_\nu$ that are marginalized over the dark energy equation of state at late times.

This paper is organized as follows. Section~\ref{sec:models} defines the dark energy models and parametrizations adopted in this work. Section~\ref{sec:datasets} describes the datasets used in this work and details our analysis. Section~\ref{sec:results} presents our main results and Section~\ref{sec:conclusion} summarizes our findings. We complement the results presented in the main text with three Appendices: Appendix~\ref{app:binw_full} shows the full reconstruction of the equation of state using all the principal components, Appendix~\ref{app:smooth_binw} a robustness test of the binned  $w(z)$ model using a smooth alternative, and Appendix~\ref{app:tanh_varysigma} a more detailed analysis of the transition model with a varying width parameter.

\section{Dark Energy Models}
\label{sec:models}
In this Section, we summarize the dark energy models analyzed in this work. We consider models that allow a phantom crossing of $w(z)$. The transition parametrization in Section~\ref{sec:tanh} captures the transition between two values of $w(z)$; the binned $w(z)$ parametrization in Section~\ref{sec:binw} is a more general model that is agnostic to the redshift evolution of dark energy. Alongside these phenomenological choices, in Section~\ref{sec:mke}, we also analyze the monodromic k-essence model: a scalar field with non-canonical lagrangian.

\subsection{Transition model}
\label{sec:tanh}
This model represents an equation of state that smoothly transitions between two values: $w_1$ at early times and $w_0$ at late times. The dark energy equation of state is parametrized as
\begin{equation}\label{eq:tanh}
    w(z) = w_0 + \left( \frac{w_1 - w_0}{2}\right)\left( 1 + \tanh\left(\frac{z - z_c}{\sigma}\right)\right).
\end{equation}
For high redshifts, $z - z_c \gg \sigma$, $w(z) \approx w_1$; conversely, for $z < z_c$ and $|z - z_c| \gg \sigma$, $w(z) \approx w_0$. The parameter $z_c$ is the redshift in which the transition is at its midpoint: $w(z_c) = (w_0 + w_1)/2$. $\sigma$ is a parameter that controls the redshift width of the transition: smaller values of $\sigma$ represent faster transitions, while larger values of $\sigma$ represent slower transitions. The energy density of this model is obtained via the integral
\begin{equation}\label{eq:rhoDE}
    \rho_\mathrm{DE} = \Omega_\mathrm{DE}\rho_\mathrm{cr}\exp\left(3\int_0^z\frac{1 + w(z')}{1 + z'}dz'\right),
\end{equation}
where $\rho_\mathrm{cr} = 3H_0^2/(8\pi G)$, $\Omega_\mathrm{DE} = 1 - \Omega_m - \Omega_r$ and the integral is calculated numerically. This model has four parameters: $w_0$, $w_1$, $z_c$ and $\sigma$. We choose to fix $\sigma = 0.05$ throughout our analysis: this value is sufficiently small to guarantee a fast transition within a redshift window of $\Delta z < 0.1$. For completeness, in Appendix \ref{app:tanh_varysigma} we present results for varying $\sigma$.

This model crosses the phantom barrier at $w = -1$ if $w_0$ and $w_1$ are chosen such that $(1 + w_0)(1 + w_1) < 0$. To avoid instabilities, we use the Parametrized Post-Friedmann (PPF) approach to calculate the dark energy perturbations \cite{ppf_cmb, ppf_de, ppf_mg}.

\subsection{\boldmath Binned \texorpdfstring{$w(z)$}{}}
\label{sec:binw}
In this model, the dark energy equation of state is given by constant values inside several redshift bins with limits $\{0, z_1, z_2, ..., z_N\}$. The equation of state can be written as
\begin{equation}\label{eq:binw}
    w(z) = \begin{cases}
        w_0, \text{ if } z < z_1; \\
        w_1, \text{ if } z_1 \leq z < z_2; \\
        w_2, \text{ if } z_2 \leq z < z_3; \\
        ... \\
        w_{N-1}, \text{ if } z_{N-1} \leq z < z_N; \\
        -1, \text{ if } z > z_N. \\
    \end{cases}
\end{equation}
At redshifts larger than the last bin, we set $w = -1$. This model can approximate any smooth dark energy model that varies slowly inside each redshift bin. Increasing the number of redshift bins improves the generality of this parametrization by making the bins narrower. However, adding many extra parameters to the model makes the analysis more computationally expensive. We focus on the dark energy behavior at very low redshifts $z < 1$, when dark energy becomes dominant in the background energy density and many supernovae are detected. 

Thus, we choose two strategies of redshift binning, with 5 and 10 bins, respectively, with constant redshift intervals:
\begin{itemize}
    \item $z_i = 0.3 \times i$, $i = 1, ..., 5$;
    \item $z_i = 0.1 \times i$, $i = 1, ..., 10$.
\end{itemize}
Even though there are SN and BAO distance measurements at higher redshifts, we restrained ourselves to the low redshift windows of $z < 1$ and $z < 1.5$. For $z > 2$, dark energy comprises less than $10\%$ of the Universe's energy density, having a reduced impact on the expansion history. Furthermore, there are fewer supernovae measurements for redshifts $z > 1$ and no DESY5 supernovae for $z > 1.13$, degrading the dark energy constraints above those redshifts. In addition, it is claimed by \cite{desi_2024_bao} that the DESI measurements that cause the preference for dynamical dark energy are those inferred from the Luminous Red Galaxy samples at redshifts $0.6 < z < 0.8$. As previous analyses from literature (e.g., \cite{desi_crossing}) suggest a phantom crossing at redshifts $z \approx 0.5$, we focus on the dynamics close to this redshift.

By imposing the continuity of the dark energy density in the edge of two consecutive redshift bins, we obtain the following expression for the evolution of the energy density 
\begin{equation}
    \rho_\mathrm{DE}(z_{i}<z<z_{i+1}) = \Omega_\mathrm{DE}\rho_\mathrm{cr}\left(\prod_{j=1}^i (1+z_j)^{3(w_{j-1}-w_j)}\right)(1+z)^{3(1+w_i)}.
\end{equation}
As before, we use the PPF framework to compute the dark energy perturbations for this model.
For robustness, we have implemented a continuous version of the discontinuous binned $w(z)$ given in Equation \eqref{eq:binw}. In Appendix \ref{app:smooth_binw}, we compare results between the discontinuous and continuous versions of the model, assessing differences in parameter constraints and goodness-of-fit. We conclude that the discontinuity in $w(z)$ does not affect our analysis.

\subsection{Monodromic k-essence}
\label{sec:mke}
\begin{figure}
    \centering
    \includegraphics[width=0.9\textwidth]{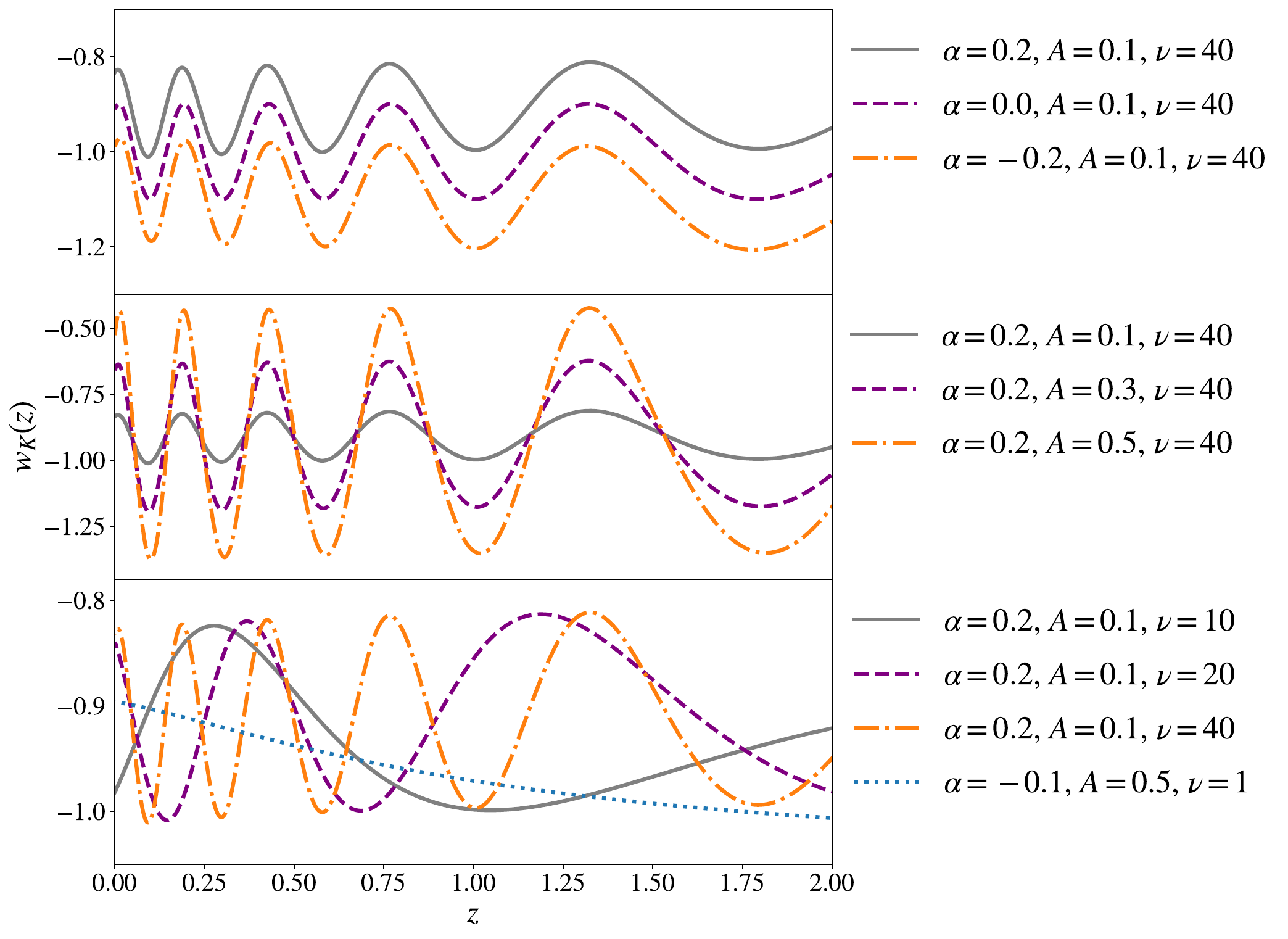}
    \caption{Evolution of the monodromic k-essence equation of state $w_K$ for selected potential parameter values. The top row shows variations of the $\alpha$ parameter, the middle row shows variations of $A$ and the bottom row shows variations of $\nu$ (in units of $M_\mathrm{pl}^{-1}$).}
    \label{fig:mke-cases}
\end{figure}

K-essence is a scalar field, minimally coupled to gravity, differing from quintessence by having a non-canonical lagrangian \cite{kessence, kessence_essentials, kessence_and_quintessence, kessence_new_view, dynamics_dark_energy}. Assuming the flat Friedmann-Robertson-Walker metric with signature $(-, +, +, +)$, a general scalar field action can be written as
\begin{equation}
    S = \int d^4x \sqrt{-g}P(X, \phi),
\end{equation}
where $X$ is the kinetic term of the scalar field given by
\begin{equation}
    X = -\frac{1}{2}\nabla_\mu\phi\nabla^\mu\phi.
\end{equation}
Non-canonical kinetically-driven fields \cite{kessence}, in which $P(X, \phi)$ is of the form
\begin{equation}\label{eq:mke_lagrangian}
    P(X, \phi) = V(\phi)(-X + X^2),
\end{equation}
can exhibit tracking behavior and scaling solutions as in some quintessence models but are not restricted to an equation of state $w \geq -1$. We use the monodromic k-essence (MKE) model introduced in \cite{mke_model}, where the function $V(\phi)$ is given by
\begin{equation}
    V(\phi) = C\left(\frac{\phi}{\phi_0}\right)^{-\alpha}(1 - A\sin(\nu\phi)).
\end{equation}
The energy density and equation of state of the field are given by
\begin{subequations}
    \begin{align}
        \rho_K &= 2X\frac{\partial P}{\partial X} - P = V(\phi)X(-1 + 3X), \\
        w_K &= \frac{P(X, \phi)}{\rho_K} = \frac{1 - X}{1 - 3X}.
    \end{align}
\end{subequations}
To find the field evolution in time, we must integrate a second-order differential equation for $\phi$, a generalization of the Klein-Gordon equation for quintessence. Considering the specific form of the lagrangian $P(X, \phi)$ in Equation~\ref{eq:mke_lagrangian}, the equation of motion is
\begin{equation}
    \dot{X} = \frac{-V'(\phi)\sqrt{2X}(-X + 3X^2) - 6HV(\phi)X(-1 + 2X)}{V(\phi)(6X - 1)},
\end{equation}
where overdots indicate derivatives with respect to cosmic time $t$ and $H$ is the Hubble factor. Our implementation solves the equation above numerically for $X$, alongside the equation for the homogeneous background field $\phi$
\begin{equation}
    \dot{\phi} = \sqrt{2X}.
\end{equation}
Like quintessence, this scalar field model exhibits tracking behavior: many initial conditions quickly converge to an attractor that determines the field evolution \cite{kessence_tracking}. This makes today's dark energy behavior highly insensitive to the field's initial conditions. Thus, we choose initial conditions according to the tracking solution \cite{mke_model}. We adjust the parameter $C$ to ensure the closure relation today, $\sum_\text{all species}\rho_i/\rho_\mathrm{cr} = 1$, using the secant method. The remaining model parameters are $\alpha$, $A$, and $\nu$, as $\phi_0$ is fixed to $1\, M_\mathrm{pl}$ without loss of generality. The MKE equation of state $w_K$ is approximately constant at high redshifts; when dark energy starts dominating the background density, $w_K$ begins oscillating. Figure~\ref{fig:mke-cases} shows evolutions of $w_K$ varying each parameter $\alpha$, $A$ and $\nu$ individually. The parameter $\alpha$ controls the mean value of $w_K$, denoted by $\Bar{w}_K$, which is approximated by
\begin{equation}
    \Bar{w}_K = -1 + \frac{\alpha}{2}.
    \label{eq:average_w_mke}
\end{equation}
While the original reference \cite{mke_model} restricted its description for $\alpha > 0$, we also consider the cases where $\alpha < 0$, in which the function $V(\phi)$ has a positive power law. This makes the equation of state $w_K$ oscillate around $\Bar{w}_K < -1$, which is relevant in light of the DESI results. The parameter $A$ controls the amplitude of the oscillations, and $\nu$ controls the frequency of the oscillations. Frequency values $\nu \lesssim 1\,M_\mathrm{pl}^{-1}$ are too small such that $w_K(z)$ does not perform a complete oscillation, providing a monotonic behavior as shown in the bottom panel of Figure~\ref{fig:mke-cases}.

The perturbations of the k-essence field $\delta\phi$ in the synchronous gauge are governed by the equation \cite{kessence_perturbations}
\begin{equation}
    A\delta\phi'' + B\delta\phi' + C\delta\phi + D = 0,
\end{equation}
where the primes $'$ denote derivatives with respect to conformal time $d\tau = dt/a$, and the coefficients are
\begin{subequations}
    \begin{align}
        A &= P_{,XX}\frac{\phi'^2}{a^2} + P_{,X}, \\
        B &= 2\mathcal{H}P_{,X} + P_{,XXX}\frac{X'\phi'^2}{2} + P_{,XX\phi}\frac{\phi'^3}{a^2} + P_{,XX}\left(3\frac{\phi'\phi''}{a^2} - \frac{\mathcal{H}\phi'^2}{a^2} \right) + P_{,X\phi}\phi', \\
        C &= -a^2P_{,\phi\phi} + k^2P_{,X} + P_{,XX\phi}X'\phi' + P_{,X\phi}\phi'' + P_{,X\phi\phi}\phi'^2 + 2\mathcal{H}P_{,X\phi}\phi', \\
        D &= 3P_{,X}h\phi',
    \end{align}
\end{subequations}
where the commas represent partial derivatives (i.e. $P_{,X} = \partial P / \partial X$), $\mathcal{H} = a'/a$ is the conformal Hubble factor and $h$ is the synchronous gauge curvature perturbation. The perturbed energy-momentum tensor components in synchronous gauge can be written as
\begin{subequations}
    \begin{align}
        \delta\rho_K = P_{,X}\delta X - P_{,\phi}\delta\phi + 2XP_{,XX}\delta X + 2XP_{,X\phi}\delta\phi,\\
        \delta P_K = P_{,X}\delta X - P_{,\phi}\delta\phi,\\
        (\rho_K + P_K)v_K = P_{,X}k\phi'\delta\phi/a^2,
    \end{align}
\end{subequations}
where $\delta X = \phi' \delta\phi' / a^2$. While we are interested in its background properties, this model is also interesting for having a sound speed $c_s^2 \approx 0$, which can induce dark energy perturbations and impact structure formation significantly. We include the k-essence perturbations in our numerical implementation of the model. While a phantom crossing for a general scalar field lagrangian would evoke instabilities in its perturbations \cite{quintom, can_de_evolve_phantom}, these instabilities are avoided by the fact that the effective sound speed $c_s^2$ vanishes at the phantom crossing \cite{eft_quintessence}.

\section{Datasets and Analysis Methodology}
\label{sec:datasets}
This Section presents the datasets used to investigate the dark energy models described in Section \ref{sec:models}. We present the individual datasets in Sections~\ref{sec:desi} to Section~\ref{sec:actlens}. Section~\ref{sec:dcs} explains the dataset combinations we considered to analyze. In Section~\ref{sec:analysis}, we detail our analysis methodology.

\subsection{DESI 2024 BAO}
\label{sec:desi}
The Dark Energy Spectroscopic Instrument has measured the baryonic acoustic oscillation signal from galaxy clustering correlations using three different tracers of matter: galaxies, quasars \cite{desi_galaxies_qso} and the Lyman-$\alpha$ forest \cite{desi_lya}. The tracers are spread over a redshift range $0.1 < z < 4.2$, divided into seven bins. Measuring the 2-point correlation function of tracer positions, one can identify the sound horizon at the baryon drag epoch, $z_d$, defined as
\begin{equation}
    r_d = \int_{z_d}^\infty \frac{c_s(z)}{H(z)}dz,
\end{equation}
where $c_s(z)$ is the baryon-photon fluid sound speed. This physical scale can be used as a ruler to measure the transverse comoving distance to the tracers at each redshift bin, which, in a flat Universe, is given by
\begin{equation}
    D_M(z) = c\int_0^z \frac{dz'}{H(z')},
\end{equation}
and the Hubble factor at the tracer redshift, which can be mapped to a distance measurement via
\begin{equation}
    D_H(z) = \frac{c}{H(z)}.
\end{equation}
The BAO measurements constrain the quantities $D_M(z)/r_d$ and $D_H(z)/r_d$. Throughout the rest of this work, we refer to this dataset simply as DESI.

\subsection{Type Ia Supernovae Catalogs}
\label{sec:sne}
Type Ia supernovae catalogs provide luminosity distance measurements, given by
\begin{equation}
    D_L(z) = c(1+z)\int_0^z \frac{dz'}{H(z')}.
\end{equation}
In this work, we use four different catalogs. The Pantheon light-curve sample is a collection of 1048 Type Ia supernovae, covering the redshift range $0.01 < z < 2.3$ \cite{pantheon_sample}. The Pantheon+ catalog comprises 1701 light curves measured from 1550 Type Ia supernovae over the redshift range $0.001 < z < 2.26$. The Union3 catalog consists of 2087 Type Ia supernovae over the redshift range $0.01 < z < 2.26$, and the public data consists of binned distance measurements \cite{union3_sn}. The Dark Energy Survey Year 5 (DESY5) catalog consists of 1635 Type Ia supernovae over the redshift range $0.1 < z < 1.13$ \cite{desy5_cosmo}. We will refer to these supernovae catalogs for conciseness as Pan, Pan+, Uni3, and DESY5, respectively.

\subsection{Planck 2018 CMB}
\label{sec:planck}
We use the Planck 2018 CMB temperature and polarization angular power spectra, including the \texttt{plik} likelihood for the TT, TE and EE spectra at multipoles $ 30 \leq \ell \leq 2508$, and the low multipole likelihoods \texttt{Commander} and \texttt{Simall} for TT and EE spectra at multipoles $\ell < 30$ \cite{planck_plik_like}. For conciseness, we will refer to this dataset as P18.

\subsection{ACT DR4 CMB Temperature and Polarization Anisotropies}
\label{sec:actdr4}
We use the ACT TT, TE, and EE power spectra from their Data Release 4, up to multipoles $\ell = 4000$. Due to the lack of large-scale polarization measurements, ACT CMB data alone cannot constrain the reionization optical depth $\tau$. Following \cite{act_ede}, we supplement ACT data with Planck 2018 power spectra up to multipole $\ell = 650$. We will refer to this combination of ACT and Planck CMB data as ACT+P650.

\subsection{ACT DR6 CMB Lensing Power Spectrum}
\label{sec:actlens}
The CMB lensing power spectrum, $\hat{C}_\ell^{\phi\phi}$, with multipoles range of $40 < \ell < 763$ in 10 bandpowers, are reported by the Atacama Cosmology Telescope (ACT) Data Release 6 (DR6) \cite{act_dr6_lensing_cp,act_dr6_lensing_measure}. ACT is a ground-based CMB survey that minimizes foreground and noise using a new pipeline, and from a lensing mass map covering $9400~\text{deg}^2$ of the sky, they report measurements of the CMB lensing amplitude with a precision of 2.3\%. 

The theoretical prediction for the CMB lensing power spectrum, $C_\ell^{\phi\phi}$, is given by the line-of-sight integral projected onto the non-linear power spectrum $P_\mathrm{NL}$ \cite{act_dr6_lensing_measure}
\begin{equation}
    \label{theory_lensing_ps}
    C_\ell^{\phi\phi} = \int dz \frac{H}{c\chi(z)^2}[W^\kappa(z)]^2 P_\mathrm{NL}(\ell/\chi(z), z)
\end{equation}
where $W^\kappa(z)$ is the lensing kernel
\begin{align}
    W^\kappa(z) = \frac{3}{2}\Omega_mH_0^2\frac{1+z}{H(z)}\frac{\chi(z)}{c}\left[\frac{\chi(z_\star)-\chi(z)}{\chi(z_\star)}\right].
\end{align}
Since massive neutrinos suppress the matter power spectrum on scales smaller than the free-streaming, and dark energy also impacts the structure formation, CMB lensing is a good source of information for these components. We will refer to the ACT DR6 lensing dataset as ACTL.

\subsection{Dataset Combinations}
\label{sec:dcs}
Among the individual datasets, several combinations can be analyzed. BAO measurements alone have a degeneracy between $H_0$ and $r_d$, which can be broken by the inclusion of CMB data. Supernovae distance measurements add extra geometric information, constraining the dark energy behavior at low redshifts. Thus, we consider combinations of CMB, BAO, and SN datasets. To assess the differences in parameter constraints among the multiple supernovae datasets, we combine each SN catalog with Planck 2018 CMB data and DESI BAO data, forming four different dataset combinations:
\begin{itemize}
    \item Planck 2018 (P18) + DESI-Y1 BAO (DESI) + Pantheon (Pan);
    \item Planck 2018 (P18) + DESI-Y1 BAO (DESI) + Pantheon + (Pan+);
    \item Planck 2018 (P18) + DESI-Y1 BAO (DESI) + Union-3 (Uni3);
    \item Planck 2018 (P18) + DESI-Y1 BAO (DESI) + DES-Y5 (DESY5).
\end{itemize}
We also assess neutrino mass constraints changing the CMB dataset from Planck 2018 to ACT+P650, as described in Section~\ref{sec:actdr4}:
\begin{itemize}
    \item ACT DR4 (ACT) + Planck 2018 $\ell < 650$ (P650) + DESI + Pan;
    \item ACT DR4 (ACT) + Planck 2018 $\ell < 650$ (P650) + DESI + Pan+;
    \item ACT DR4 (ACT) + Planck 2018 $\ell < 650$ (P650) + DESI + Uni3;
    \item ACT DR4 (ACT) + Planck 2018 $\ell < 650$ (P650) + DESI + DESY5.
\end{itemize}
As a probe of large-scale structure, CMB lensing provides tight constraints on the neutrino mass sum. We also include ACT DR6 CMB Lensing (ACTL) data with the ACT+P650 anisotropy data:
\begin{itemize}
    \item ACT + P650 + ACT DR6 Lensing (ACTL) + DESI + Pan;
    \item ACT + P650 + ACT DR6 Lensing (ACTL) + DESI + Pan+;
    \item ACT + P650 + ACT DR6 Lensing (ACTL) + DESI + Uni3;
    \item ACT + P650 + ACT DR6 Lensing (ACTL) + DESI + DESY5.
\end{itemize}

We neglect correlations between datasets; thus, the total likelihood is simply the product of each likelihood. The different supernovae catalogs cannot be trivially combined since they significantly overlap in supernovae.

\subsection{Analysis Methodology}
\label{sec:analysis}
The three dark energy models discussed in Section \ref{sec:models} were implemented in a modified version of \texttt{CAMB}\footnote{\href{https://github.com/cmbant/CAMB}{https://github.com/cmbant/CAMB}} \cite{camb_original} and are publicly available. The phenomenological models (transition and binned $w(z)$ models) are in the same code\footnote{\href{https://github.com/SBU-COSMOLIKE/CAMBLateDE}{https://github.com/SBU-COSMOLIKE/CAMBLateDE}}, while the scalar field monodromic k-essence model is implemented in a different one\footnote{\href{https://github.com/SBU-COSMOLIKE/CAMB-Monodromic}{https://github.com/SBU-COSMOLIKE/CAMB-Monodromic}}. 

For every dataset combination and each dark energy model, we sample over the parameter posterior distribution with Markov Chain Monte Carlo (MCMC) methods, specifically using the Metropolis-Hastings algorithm implemented in the \texttt{Cobaya}\footnote{\href{https://github.com/CobayaSampler/cobaya}{https://github.com/CobayaSampler/cobaya}} code \cite{cobaya_code, cosmo_mcmc, fast_slow}. We use \texttt{CoCoA}\footnote{\href{https://github.com/CosmoLike/cocoa}{https://github.com/CosmoLike/cocoa}}, the Cobaya-Cosmolike Joint Architecture \cite{cosmolike} framework, to run our two modified \texttt{CAMB}s with \texttt{Cobaya}. We make use of \texttt{GetDist}\footnote{\href{https://github.com/cmbant/getdist}{https://github.com/cmbant/getdist}} \cite{getdist} to perform statistical analysis of the MCMC samples and produce the figures shown in this work.

\begin{table}
    \centering
    \begin{tabular}{|c|c|c|}
    \hline
    Model & Parameters & Priors \\
    \hline
    $\Lambda$CDM & $\Omega_\text{c}h^2$& $\mathcal{U}[0.001, 0.99]$\\
     & $\Omega_\text{b}h^2$& $\mathcal{U}[0.005, 0.1]$\\
     & $\ln(10^{10}A_\text{s})$& $\mathcal{U}[1.61, 3.91]$\\
     & $n_\text{s}$& $\mathcal{U}[0.8, 1.2]$\\
     & $\tau$& $\mathcal{U}[0.01, 0.8]$\\
     & $100\times\theta_*$& $\mathcal{U}[0.5, 10]$\\
     & $\sum m_\nu$ (eV) & $\mathcal{F}[0.06]$ or $\mathcal{U}[0.01, 0.6]$\\
    \hline
      $w_0w_a$ &$w_0$  & $\mathcal{U}[-3, -0.01]$ \\
       &$w_a$ &  $\mathcal{U}[-3, 1]$\\
    \hline
      Transition &$w_0$  & $\mathcal{U}[-3, -0.01]$ \\
       &$w_1$ &  $\mathcal{U}[-3, 1]$\\
       &$z_c$ & $\mathcal{U}[0.2, 1.0]$ \\
       &$\sigma$ & $\mathcal{F}[0.05]$\\
    \hline
      Binned $w(z)$ &$w_0$ & $\mathcal{U}[-3, -0.01]$ \\
      &$w_{i>0}$ & $\mathcal{U}[-3, 1]$ \\
    \hline
      k-essence &$\alpha$ & $\mathcal{U}[0.01, 3]$ \\
      &$A$ & $\mathcal{U}[0.01, 0.9]$ \\
      &$\nu$ ($M_\mathrm{pl}^{-1}$) & $\mathcal{U}[\nu_i - 2, \nu_i + 2]$, $i = 1, 2, 3, 4$ {$^a$} \\
    \hline
    \end{tabular}\\
    \footnotesize{$^a$ See Section~\ref{sec:results_mke} for the values for $\nu_1$, $\nu_2$, $\nu_3$ and $\nu_4$.}\\
    \caption{Priors for $\Lambda$CDM and dark energy parameters. $\mathcal{U}[a, b]$ represents an uniform distribution from $a$ to $b$, whereas $\mathcal{F}[x]$ represents a fixed value.}
    \label{tab:priors}
\end{table}

The MCMCs are performed in two scenarios: fixing $\sum m_\nu = 0.06\mathrm{~eV}$ or varying $\sum m_\nu$. We use a single massive neutrino and two massless neutrinos in the former case. In the latter, we adopt three degenerate massive neutrinos to avoid the choice of a given mass hierarchy \cite{DEG_model, nu_status_ca_constraints}. We explore the dark energy parameter posteriors, as well as the six cosmological parameters $\{ \theta_*, \Omega_ch^2, \Omega_bh^2, A_s, n_s, \tau \}$ with/without varying $\sum m_\nu$ according to priors shown in Table~\ref{tab:priors}. We choose to sample over $\theta_*$, the angular scale of the sound horizon at recombination, instead of $H_0$. Since CMB data puts very stringent constraints on $\theta_*$, this choice optimizes the number of accepted points during the MCMC.
We also run MCMCs considering the cosmological constant and $w_0w_a$ parametrization models for dark energy. We use the Casarini prescription \cite{casarini1, casarini2} for including nonlinear effects in the matter power spectrum. We use the Gelman-Rubin criterion to assess whether the chains converged, stopping when $|R-1| < 0.01$. For datasets with ACT DR4 or CMB lensing from ACT DR6, we use increased CAMB accuracy settings: \texttt{lmax = 7000}, \texttt{lens\_potential\_accuracy = 4}, \texttt{lens\_margin = 1250}, \texttt{lSampleBoost = 1}, and \texttt{lAccuracyBoost = 1} as recommended by \cite{act_dr6_lensing_cp}.

To compare how well each dark energy model fits the datasets, we maximize the likelihood and find the minimum value of $\chi^2$ starting from the maximum a posteriori sample obtained in the MCMCs. Recent works~\cite{procoli, pinc, prospect} have shown that classical minimizers, such as \texttt{Py-BOBYQA}, fail to reach the global minimium of $\chi^2$, getting stuck in local minima. Simulated annealing, a stochastic minimization heuristic, has been proposed as a means to reach global minima of cosmological likelihood functions, outperforming deterministic minimizers. We have implemented a simple simulated annealing optimizer in \texttt{Cobaya} and we use it to find the minimum values of $\chi^2$ and assess the goodness-of-fit of each model.

During our initial MCMC runs, we observed that the $\nu$ parameter, representing the oscillatory frequency in the MKE model, has a multimodal posterior distribution. The Metropolis-Hastings algorithm cannot explore the $\nu$ modes efficiently. To avoid this issue, one can employ a nested sample algorithm such as \texttt{PolyChord} \cite{polychord1,polychord2} or perform a minimization procedure; we opted for the latter. We minimize the total $\chi^2$ with fixed values of $\nu \in [0.1, 1, 2, ..., 50]$, finding the four frequency values with lowest $\chi^2_\text{tot}$. Then, we run MCMCs varying $\nu$ by $\pm 2 M_\mathrm{pl}^{-1}$ around each of those values. We validate our pipeline by reproducing the DESI results for the $w_0w_a$ parametrization.

\section{Results}
\label{sec:results}
This Section presents the results of the analysis described in Section~\ref{sec:datasets}. This Section is organized as follows: \ref{sec:constraints} presents parameter constraints obtained through the parameter sampling for each dark energy model; \ref{sec:chi2} presents goodness-of-fit results for all models, performing a model comparison; \ref{sec:mnu} presents constraints on the neutrino mass sum.

\subsection{Model Parameter Constraints}
\label{sec:constraints}

\subsubsection{Transition model}

\begin{figure}[h]
    \centering
    \includegraphics[width=0.85\textwidth]{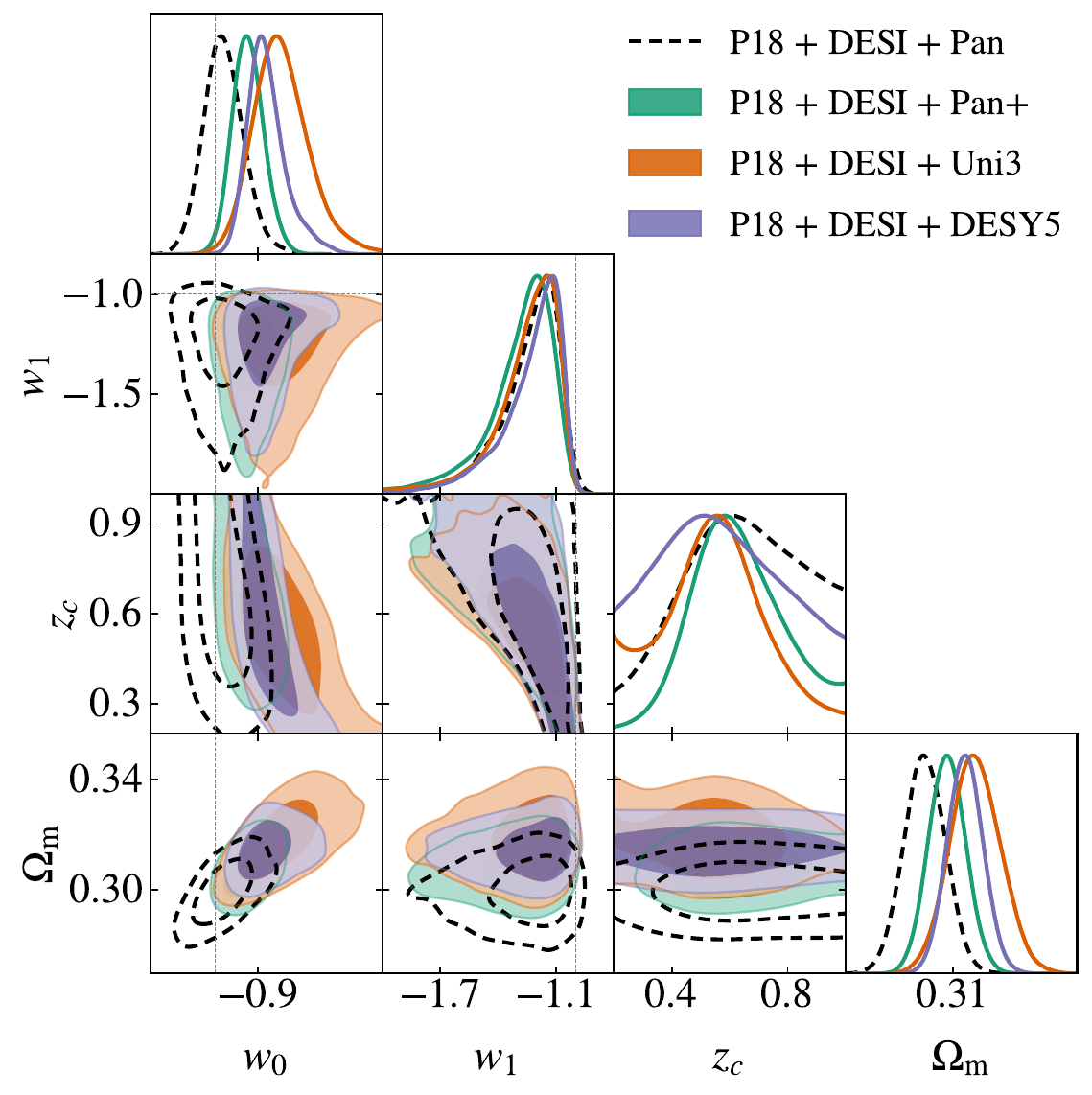}
    \caption{2D confidence contours ($68\%$ and $95\%$) of dark energy parameters $w_0$, $w_1$ and $z_c$ from the transition parametrization in Equation~\ref{eq:tanh}, as well as $\Omega_m$.}
    \label{fig:tanh_triangle}
\end{figure}

Figure~\ref{fig:tanh_triangle} shows confidence contours for the transition model described in Section~\ref{sec:tanh}. For Pantheon+, Union3, and DESY5, the $95\%$ confidence region is contained in the quadrant of the $w_0-w_1$ plane where $w_0 > -1$ and $w_1 < -1$, suggesting a transition between a phantom regime to a non-phantom behavior. Recent works \cite{steinhardt_de} show that quintessence models that satisfy $w \geq -1$ can mimic the phantom crossing $w_0w_a$ models preferred by DESI BAO. Our result corroborates the mild preference for phantom crossing using an alternative parametrization that directly probes the equation of state via $w_0$ and $w_1$. Results considering Pantheon SN are compatible with $w_0 = -1$ but also show a mild preference for $w_1 < -1$. The difference in $w_0$ constraints between the different SN datasets are caused by the correlation between $w_0$ and $\Omega_m$ observed in the lower left panel of Figure~\ref{fig:tanh_triangle}. As with the $w_0w_a$ parametrization, SN datasets that prefer higher $\Omega_m$ values will also prefer higher $w_0$ values. The transition redshift $z_c$ is weakly constrained at around $z_c \approx 0.6$. The 1D marginalized constraints (68\% limits) for the dark energy parameters are: 
\begin{itemize}
    \item \makebox[4.1cm]{P18 + DESI + Pan:\hfill} $w_0 = -0.982\pm 0.044$, $w_1 = -1.24^{+0.18}_{-0.08}$, $z_c = 0.65^{+0.25}_{-0.18}$;
    \item \makebox[4.1cm]{P18 + DESI + Pan+:\hfill} $w_0 = -0.924^{+0.034}_{-0.038}$, $w_1 = -1.30^{+0.20}_{-0.09}$, $z_c = 0.63^{+0.15}_{-0.18}$;
    \item \makebox[4.1cm]{P18 + DESI + Uni3:\hfill} $w_0 = -0.848^{+0.053}_{-0.068}$, $w_1 = -1.28^{+0.21}_{-0.07}$, $z_c = 0.56\pm 0.17$;
    \item \makebox[4.1cm]{P18 + DESI + DESY5:\hfill} $w_0 = -0.878^{+0.033}_{-0.047}$, $w_1 = -1.24^{+0.19}_{-0.06}$, $z_c = 0.57^{+0.18}_{-0.27}$.
\end{itemize}

\begin{figure}[h]
    \centering
    \includegraphics[width=0.9\textwidth]{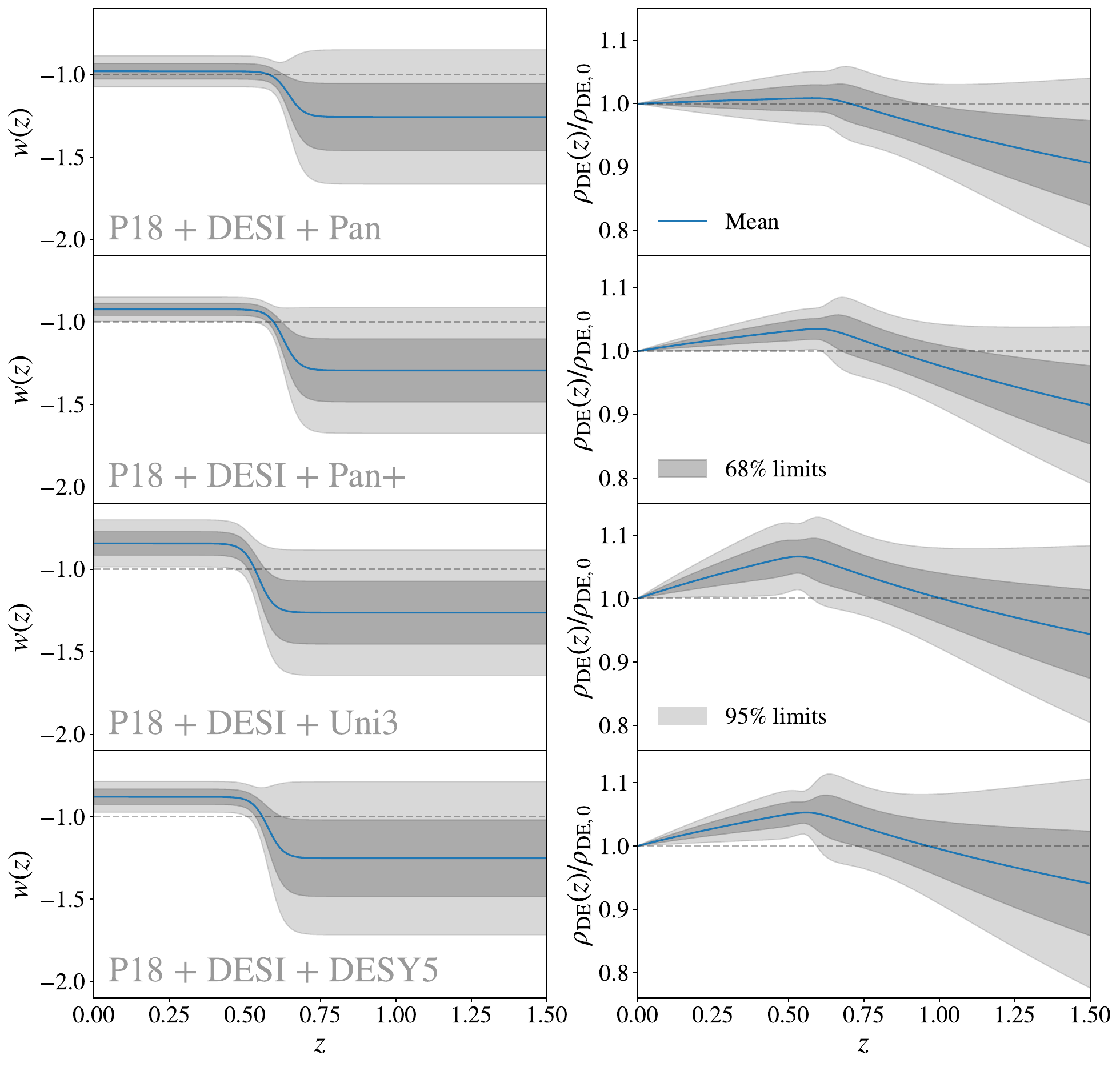}
    \caption{Reconstruction of the dark energy equation of state and density considering the transition dark energy parametrization, Equation~\ref{eq:tanh}. The blue line represents the mean model, and the darker and lighter gray bands represent the $68\%$ and $95\%$ confidence regions, respectively. The dashed black line indicates the corresponding $\Lambda$CDM values. Each row indicates a different dataset combination, all containing DESI BAO and Planck 2018 CMB data but using a different SN dataset.}
    \label{fig:tanh_reconstr}
\end{figure}

Figure~\ref{fig:tanh_reconstr} shows the constraints in the equation of state as a function of redshift, $w(z)$, and the dark energy density $\rho_\mathrm{DE}(z)$. Compared with $\Lambda$CDM, the phantom crossing behavior implies increased dark energy at redshifts $z < 0.5$ and a smaller density at higher redshifts. These results agree with previous literature analyses using the $w_0w_a$ parametrization and other reconstruction methods such as~\cite{desi_crossing}.

We have also assessed constraints with varying values of $\sigma$ with a prior $\mathcal{U}[0.01, 0.5]$, and the results are qualitatively equivalent, with wider posteriors for all parameters. The results are shown in Appendix~\ref{app:tanh_varysigma}.

We have verified by running the corresponding chains changing Planck 2018 CMB data by ACT data, as well as the inclusion of ACT CMB lensing, have no significant effect on the dark energy parameter constraints.

\subsubsection{\boldmath Binned \texorpdfstring{$w(z)$}{}}
\begin{figure}[t!]
    \centering
    \includegraphics[width=\textwidth]{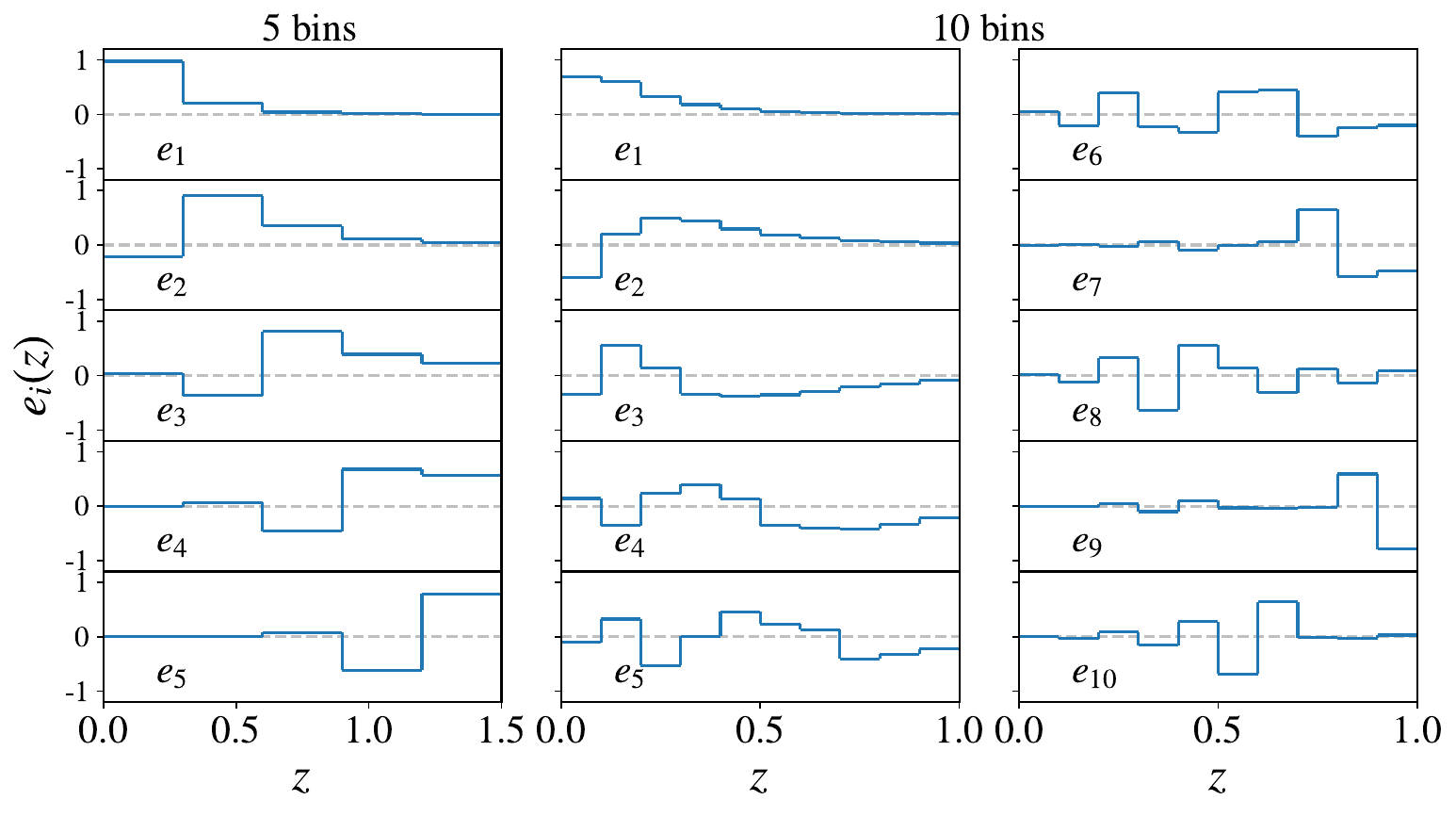}
    \caption{Principal Components ${e}_i(z)$ of the binned $w(z)$ model, defined in Equation~\ref{eq:binw}, obtained from the samples using the P18 + DESI + Pan+ dataset. The left panel shows the PCs for the case with 5 bins, while the middle and right panels show the PCs for the 10 bins case.}
    \label{fig:binw_pcs}
\end{figure}

Due to the large number of dark energy parameters and the complex correlations among them in the binned $w(z)$ model, it is difficult to visualize and interpret the constraints on the individual $w_i$'s for each redshift bin $i$. For this purpose, we perform a principal component analysis (PCA) \cite{PCA_Huterer_Starkman, PCA_Huterer_Cooray, how_many_DE_params_Linder_Huterer, DE_Experiments_with_PCA, PCA_Huterer_Peiris} of the $w_i$ samples obtained in the MCMCs. This process consists of extracting the sample covariance matrix of the $w_i$ parameters from the MCMC. Since this matrix is real and symmetric, it can be diagonalized. The eigenvectors of the sample covariance are the principal components ${e}_i$, $i = 1, ..., N_\mathrm{bins}$. Any given sample of parameters $\{w_0, ..., w_{N_\mathrm{bins}-1}\}$ represents a $w(z)$ evolution given by Equation~\ref{eq:binw} which can be rewritten in terms of the principal components as
\begin{equation}\label{eq:pca}
    w(z) = -1 + \sum_{i = 1}^{N_\mathrm{bins}}\alpha_i {e}_i(z),
\end{equation}
where we choose $\alpha = (0, ..., 0)$ to be equivalent to $w(z) = -1$\footnote{In implementations of PCA, the null vector in the principal component space maps to the mean value of the parameters in the feature space. This translation of the $\alpha_i$ parameters does not affect the eigenvectors ${e}_i$ since the covariance matrix is invariant by translations.}. Figure~\ref{fig:binw_pcs} shows the principal components ${e}_i(z)$. We order the principal components by increasing variance. We remark that, for both $N_\mathrm{bins} = 5$ and $N_\mathrm{bins} = 10$, the first principal component ${e}_1$ represents a thawing behavior, progressively deviating the dark energy equation of state from -1 in the lowest redshift bins.

\begin{figure}[h]
    \centering
    \includegraphics[width=\textwidth]{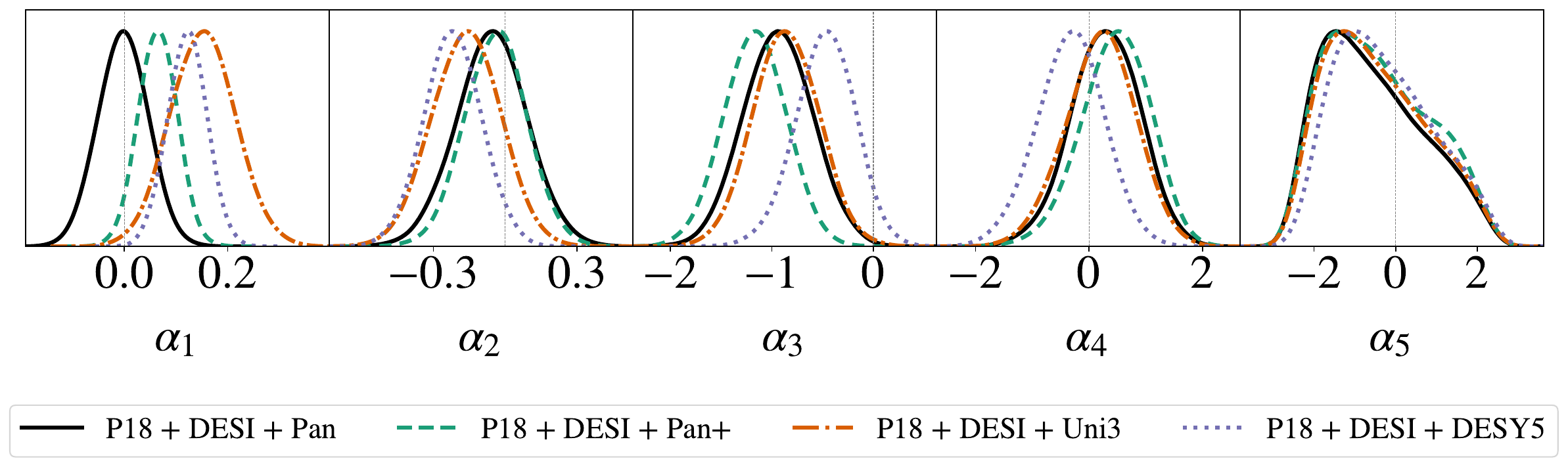}\\
    \includegraphics[width=\textwidth]{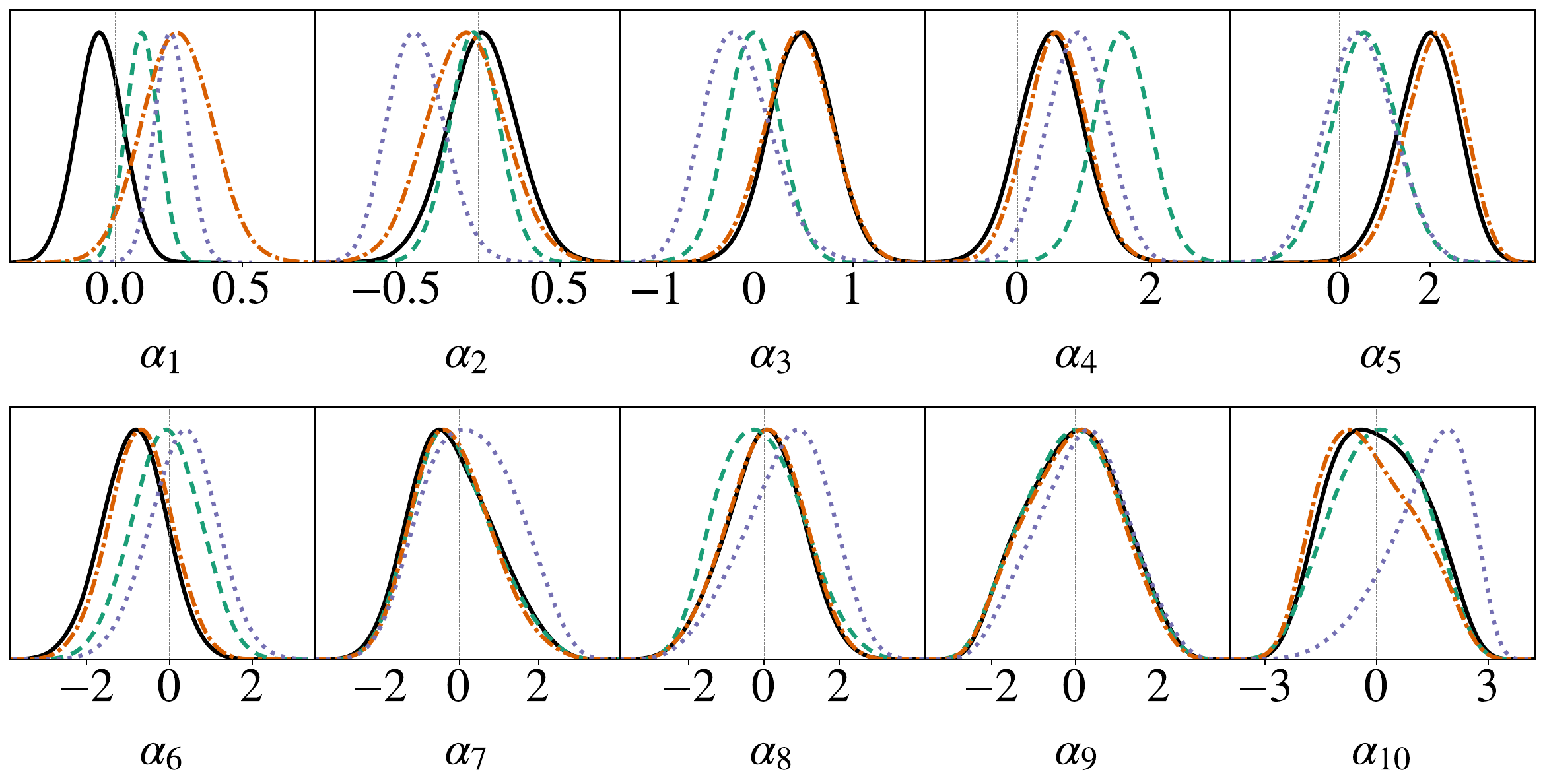}
    \caption{1D marginalized constraints on the Principal Component amplitudes $\alpha_i$. The top row shows constraints for $N_\mathrm{bins} = 5$, while the middle and bottom rows show constraints for $N_\mathrm{bins} = 10$. Each panel shows the constraints for a single PC amplitude $\alpha_i$, and each line type corresponds to a different dataset combination. At each panel, a vertical line indicates $\alpha_i = 0$. }
    \label{fig:binw_alphas}
\end{figure}

Figure~\ref{fig:binw_alphas} shows the 1D marginalized distributions for the $\alpha_i$ components for both binning choices, $N_\mathrm{bins} = 5$ and $N_\mathrm{bins} = 10$. We observe that Pantheon+, Union3, and DESY5 show a mild preference for the thawing principal component, $\alpha_1 > 0$, which is not observed in the Pantheon analysis. This is connected to previous analyses from the literature using the $w_0w_a$ parametrization \cite{desi_2024_bao, union3, desy5_cosmo} and the results from the transition model, in which all supernovae except for Pantheon show a preference for $w_0 > -1$. Reference~\cite{scalar_fields_desi}, which uses thawing quintessence models, argues that this signal is caused by a mild discrepancy in the $\Omega_m$ parameter between P18+DESI and the newer SN datasets: Pantheon SN, which prefers a value of $\Omega_m$ compatible with BAO and CMB, does not prefer thawing behavior.

For $N_\mathrm{bins} = 5$, we also observe a preference for a negative amplitude of the ${e}_3$ principal component, $\alpha_3 < 0$, and other components are compatible with zero amplitude. The 1D marginalized constraints (68\% limits) on $\alpha_3$ are:
\begin{itemize}
    \item \makebox[4.1cm]{P18 + DESI + Pan:\hfill} $\alpha_3 = -0.94 \pm 0.35$;
    \item \makebox[4.1cm]{P18 + DESI + Pan+:\hfill} $\alpha_3 = -1.17 \pm 0.31$;
    \item \makebox[4.1cm]{P18 + DESI + Uni3:\hfill} $\alpha_3 = -0.86 \pm 0.33$;
    \item \makebox[4.1cm]{P18 + DESI + DESY5:\hfill} $\alpha_3 = -0.47^{+0.31}_{-0.28}$.
\end{itemize}

\begin{figure}[h]
    \centering
    \includegraphics[width=0.9\textwidth]{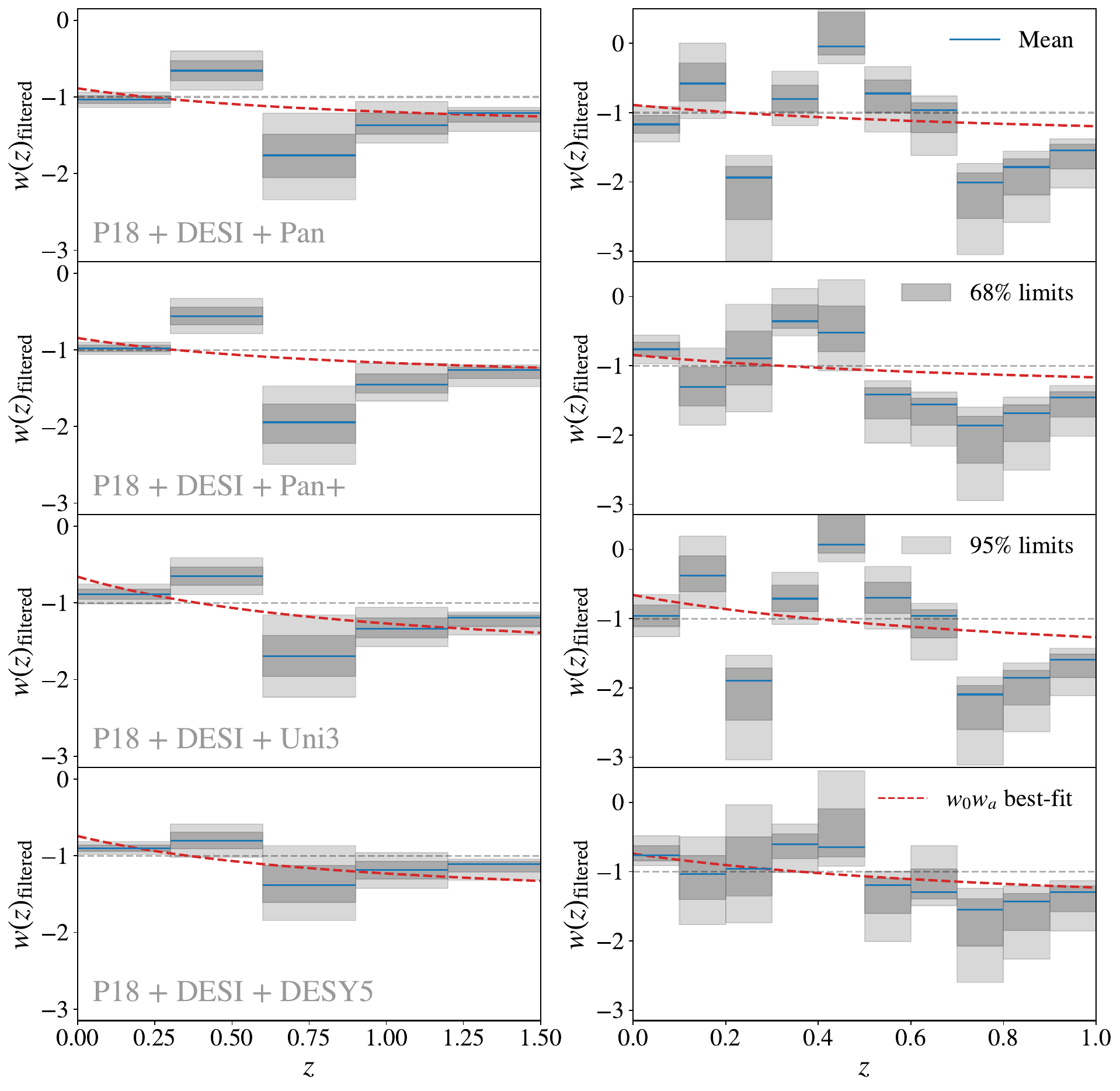}
    \caption{Filtered constraints of the dark energy equation of state using the binned $w(z)$, Equation~\ref{eq:binw}, for the different SN datasets. Here, we only show the effect of the principal components whose amplitude constraints deviate from zero: for the 5 bins, we only show the effect of ${e}_1$ and ${e}_3$, while for 10 bins we show the effect of ${e}_1$, ${e}_4$ and ${e}_5$. The mean and confidence intervals are inferred from the corresponding PC amplitudes $\alpha_i$. The blue line indicates the mean value, and the darker and lighter gray bands indicate $68\%$ and $95\%$ limits. We also include the best-fit $w_0w_a$ model as a red dashed line for comparison purposes. The left panels show the filtered constraints for $N_\mathrm{bins} = 5$, while the right panels show the filtered constraints for $N_\mathrm{bins} = 10$. Each row corresponds to a different dataset combination. We emphasize that, since we are discarding principal components amounting to noise, the error bars shown in this Figure are smaller than the actual constraints obtained from the MCMCs, and therefore there isn't necessarily a tension between the reconstructed $w(z)$ across different datasets. See Appendix~\ref{app:binw_full} for the full reconstruction of $w(z)$.}
    \label{fig:binw_filtered}
\end{figure}
With negative amplitude, the behavior of ${e}_3$, depicted in Figure~\ref{fig:binw_pcs}, increases the equation of state in the bin $0.3 < z < 0.6$ and decreases the equation of state in the subsequent bins, slowly converging to $w = -1$ at the last bins. This behavior is non-monotonic and cannot be described by a monotonic model such as the $w_0w_a$ parametrization. To visualize this non-standard $w(z)$ behavior being picked up by data, we present in the left panels of Figure~\ref{fig:binw_filtered} the filtered constraints on the dark energy equation of state, where the mean and error bars are inferred solely from $\alpha_1$ and $\alpha_3$, i.e., setting the other amplitudes to zero. We emphasize that the error bars shown in Figure~\ref{fig:binw_filtered} correspond to the uncertainties of physically motivated models only if they have vanishing projections on the noisier principal components. In Appendix~\ref{app:binw_full}, we show the full reconstruction of $w(z)$, with the full error bars on $w_i$. Although Figure \ref{fig:binw_filtered} shows noticeable differences on the filtered $w(z)$ constraints across the datasets, the full error bars indicate that there is no tension between the predicted evolution of $w(z)$ from different datasets. For Pantheon, Pantheon+, and Union3, we notice an increase of $w(z)$ in the second redshift bin $0.3 < z < 0.6$, followed by a decrease in the third bin $0.6 < z < 0.9$. Interestingly, assuming the binned $w(z)$ model, the dark energy equation of state today is consistently lower than that of the best-fit $w_0w_a$ model, with $w(z)$ peaking at the second bin. Looking forward at the 10 bins result, in the right panels of Figure~\ref{fig:binw_filtered}, we can further localize bins with increased $w(z)$ at redshifts $0.3 < z < 0.5$. The DESY5 case shows the least preference for the ${e}_3$ principal component and is consistent with a monotonic behavior.

For $N_\mathrm{bins} = 10$, apart from the thawing principal component ${e}_1$, we observe a mild preference for the two additional principal components: Pantheon+ and DESY5 show evidence for $\alpha_4 > 0$, whereas Pantheon and Union3 show evidence for ${\alpha}_5 > 0$. Finally, DESY5 also slightly prefers $\alpha_2 < 0$. The 1D marginalized constraints (68\% limits) are:
\begin{itemize}
    \item \makebox[4.1cm]{P18 + DESI + Pan:\hfill} $\alpha_4 = 0.50\pm 0.48$, $\alpha_5 = 1.94^{+0.71}_{-0.61}$;
    \item \makebox[4.1cm]{P18 + DESI + Pan+:\hfill} $\alpha_4 = 1.55\pm 0.42$, $\alpha_5 = 0.55\pm 0.70$;
    \item \makebox[4.1cm]{P18 + DESI + Uni3:\hfill} $\alpha_4 = 0.58\pm 0.45$, $\alpha_5 = 2.09^{+0.69}_{-0.60}$;
    \item \makebox[4.1cm]{P18 + DESI + DESY5:\hfill} $\alpha_4 = 0.89\pm 0.44$, $\alpha_5 = 0.45\pm 0.75$, $\alpha_2 = -0.38^{+0.16}_{-0.19}$.
\end{itemize}
The two principal components ${e}_4$ and ${e}_5$ represent oscillatory behavior with multiple phantom crossings. In the right panels of Figure~\ref{fig:binw_filtered} we show the filtered constraints on $w(z)$ inferred from $\alpha_1$, $\alpha_4$ and $\alpha_5$. We remark that we choose the same filtered principal components for all datasets. As in the case of $N_\mathrm{bins} = 5$, we also observe an increase of $w(z)$ in the redshift window $0.3 < z < 0.5$. Once again, DESY5 is the SN dataset that shows the least preference for an oscillatory behavior, consistent with the $w_0w_a$ behavior. The full constraints on $w_i$ are shown in Appendix~\ref{app:binw_full}.

In Figure~\ref{fig:chi2_vs_alpha}, we show the 2-dimensional marginalized distributions of $\chi^2_\mathrm{BAO}$ versus the amplitudes of the principal components listed above. For the 5 bins case, we observe a clear trend that $\chi^2_\mathrm{BAO}$ is minimized at around $\alpha_3 \approx -1$. For the 10 bins case, considering Pantheon+ or DESY5, $\chi^2_\mathrm{BAO}$ is minimized for $\alpha_4 \approx 1.5$, whereas for Pantheon and Union3, the minimum occurs for $\alpha_5 \approx 2$. In all cases, DESI BAO is decisive to push the amplitude constraints away from zero.

\begin{figure}
    \centering
    \includegraphics[width=0.3\linewidth]{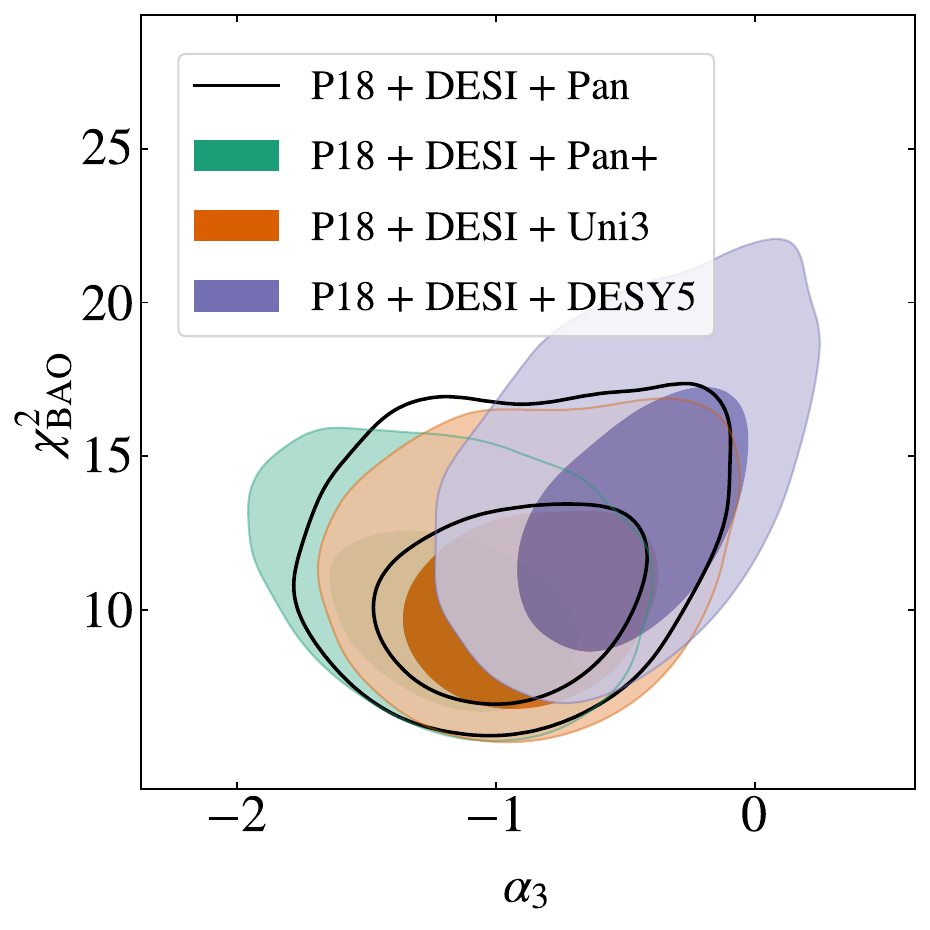}
    \includegraphics[width=0.3\linewidth]{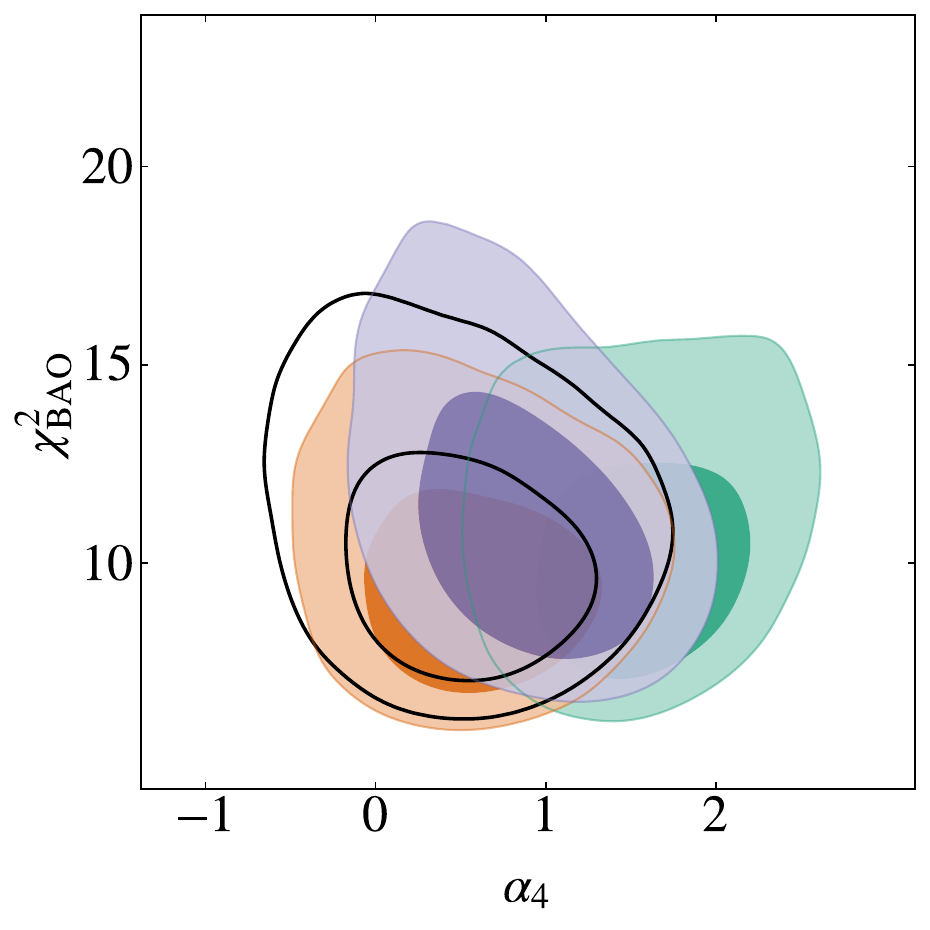}
    \includegraphics[width=0.3\linewidth]{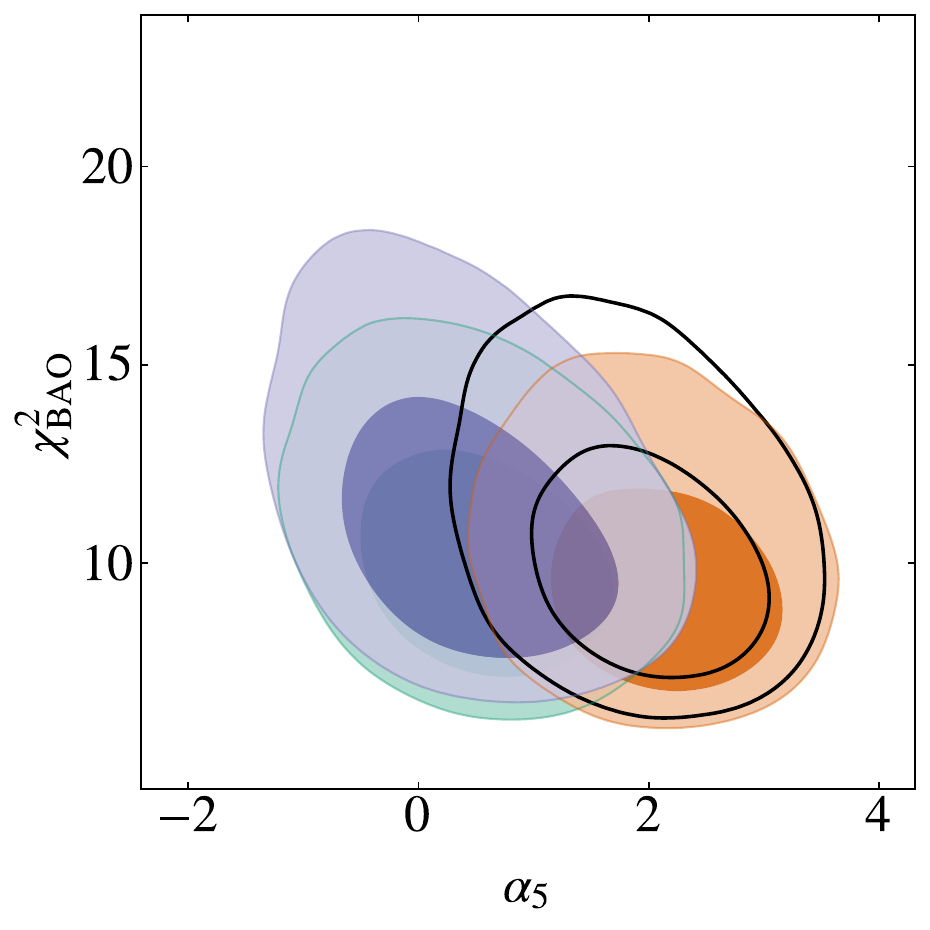}
    \caption{2-dimensional marginalized distributions of $\chi^2_\mathrm{BAO}$ and the principal component amplitudes whose mean deviates from zero. In the left panel, we show $\alpha_3$ for the 5 bins case, in the middle panel we show $\alpha_4$ for the 10 bins case and in the right panel we show $\alpha_5$, also for the 10 bins case.}
    \label{fig:chi2_vs_alpha}
\end{figure}

One may be concerned that the discontinuities in $w(z)$ plague the analysis. In Appendix~\ref{app:smooth_binw}, we present a continuous version of the binned $w(z)$ model and we show that the impact of the discontinuities in $\chi^2$ does not lead to significant differences.

\subsubsection{Monodromic k-essence}
\label{sec:results_mke}
\begin{figure}[h]
    \centering
    \includegraphics[width=0.45\textwidth]{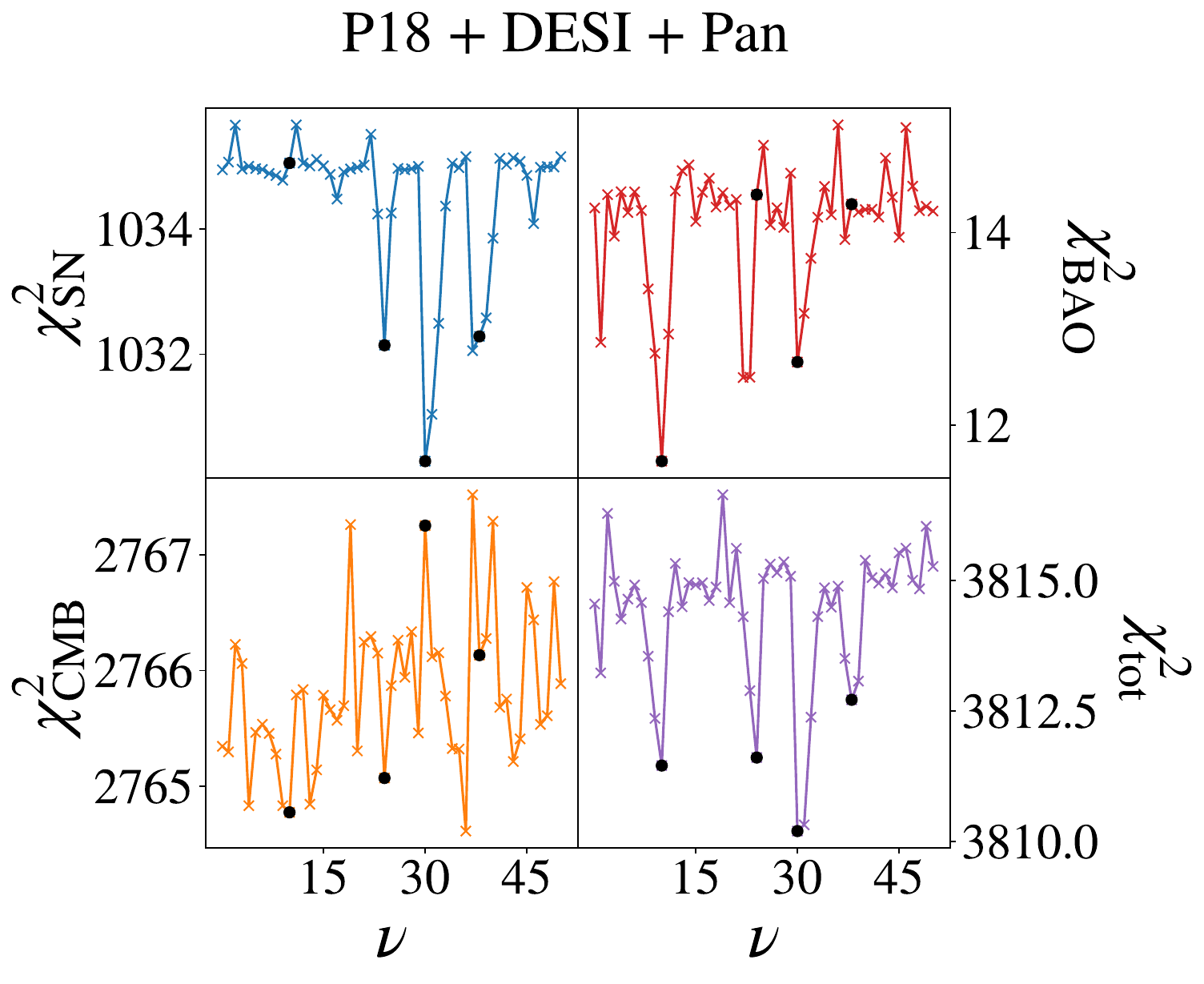}
    \includegraphics[width=0.45\textwidth]{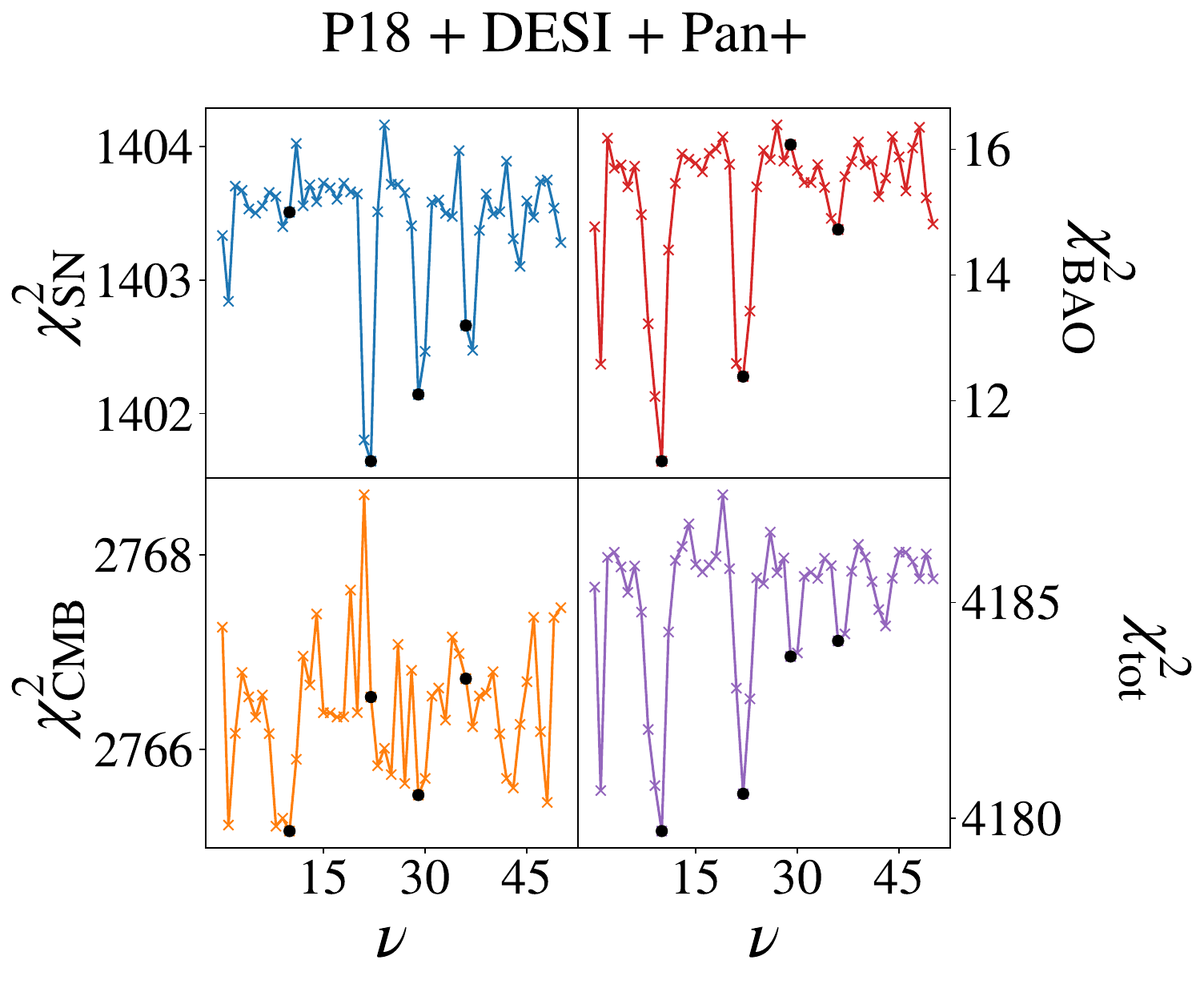}\\
    \includegraphics[width=0.45\textwidth]{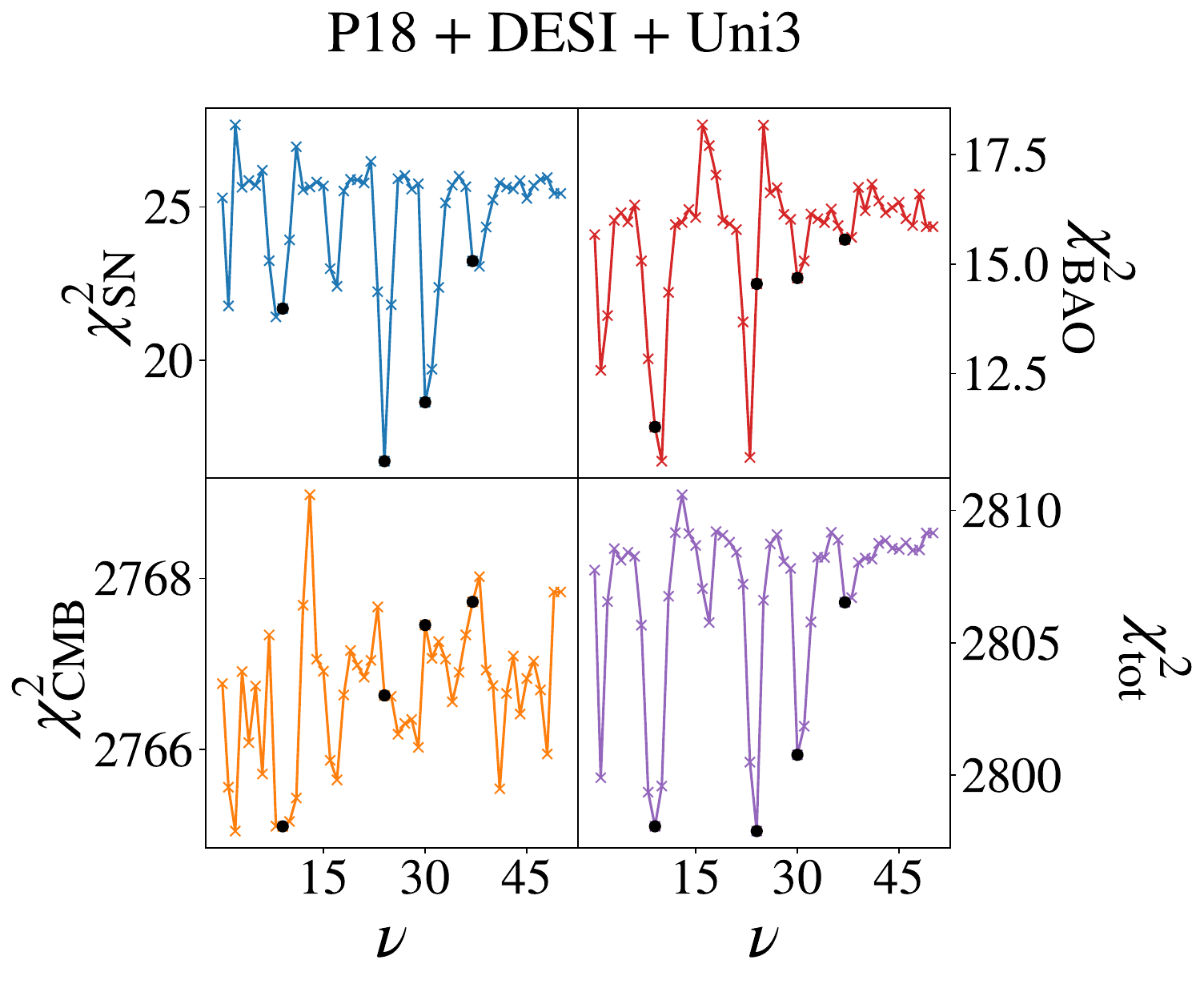}
    \includegraphics[width=0.45\textwidth]{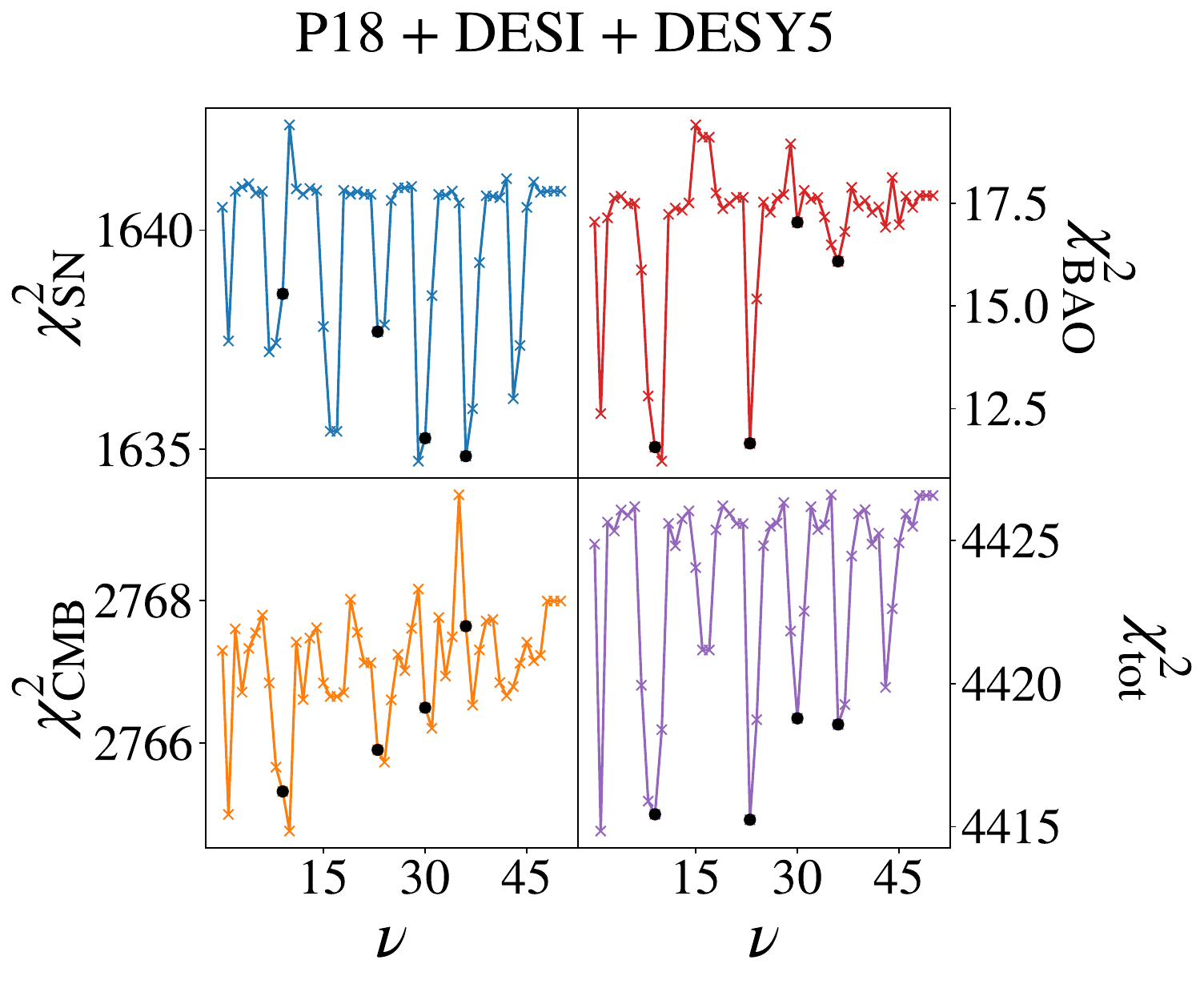}
    \caption{Minimum $\chi^2$ values for different fixed frequencies $\nu$. The black circles indicate the four frequency values that minimize $\chi^2_\mathrm{tot}$ and were chosen to run the MCMCs.}
    \label{fig:mke_freqs}
\end{figure}

Due to the multimodal posterior distribution of the parameter $\nu$, we minimize the total $\chi^2$ over all other cosmological parameters with fixed values of the frequency $\nu \in [0.1, 1, 2, 3,..., 50]$, finding the values preferred by data. The results of this minimization process are shown in Figure~\ref{fig:mke_freqs}, showing $\chi^2_\mathrm{SN}$, $\chi^2_\mathrm{BAO}$, $\chi^2_\mathrm{CMB}$ and $\chi^2_\mathrm{tot}$ for each value of $\nu$ for all datasets. Out of the tested frequency values, those that minimize $\chi^2_\mathrm{tot}$ are, in $M_\mathrm{pl}^{-1}$:
\begin{itemize}
    \item \makebox[4.1cm]{P18 + DESI + Pan:\hfill} $\nu \in  [1, \nu_1=10, \nu_2=24, \nu_3=30, \nu_4=38]$;
    \item \makebox[4.1cm]{P18 + DESI + Pan+:\hfill} $\nu \in [1, \nu_1=10, \nu_2=22, \nu_3=29, \nu_4=36]$;
    \item \makebox[4.1cm]{P18 + DESI + Uni3:\hfill} $\nu \in [1, \nu_1=9, \nu_2=24, \nu_3=30, \nu_4=37]$;
    \item \makebox[4.1cm]{P18 + DESI + DESY5:\hfill} $\nu \in [1, \nu_1=9, \nu_2=23, \nu_3=30, \nu_4=36]$.
\end{itemize}
As shown in Figure~\ref{fig:mke-cases}, with a frequency $\nu = 1 M_\mathrm{pl}^{-1}$, $w(z)$ doesn't complete a full oscillation during the cosmological history, thawing at low redshifts, a behavior similar to $w_0w_a$. We remark that, for DESY5, $\nu = 1M_\mathrm{pl}^{-1}$ is the frequency value that overall minimizes $\chi^2$. However, higher frequency values can yield lower $\chi^2$ values for other supernovae, improving the overall fit. To explore the low-frequency region of the parameter space, we run MCMCs varying the frequency with a prior of $\nu \in [0, 3]$ in $M_\mathrm{pl}^{-1}$.

It is interesting to notice that, for the different datasets, the preferred frequencies are similar, differing by at most $2 M_\mathrm{pl}^{-1}$. However, since the minima of $\chi^2$ are very narrow, this difference can be enough to make one frequency value not be preferred by another dataset: for instance, $\nu = 24 M_\mathrm{pl}^{-1}$, which minimizes $\chi^2$ under Pantheon, is not preferred by Pantheon+. We run MCMCs varying the $\nu$ parameter by $\pm 2 M_\mathrm{pl}^{-1}$ around the four preferred values of each dataset with $\nu > 3$. We label the central values them in ascending order as $\nu_1$, $\nu_2$, $\nu_3$, and $\nu_4$.

\begin{figure}
    \centering
    \includegraphics[width=0.65\linewidth]{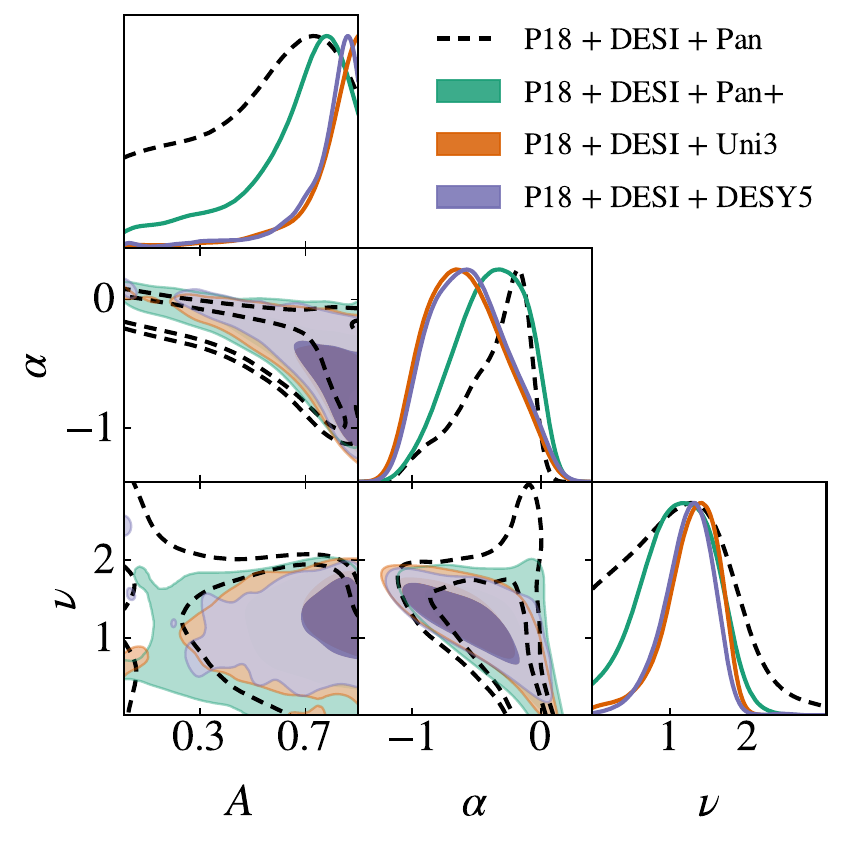}
    \caption{2D confidence contours (68\% and 95\%) of monodromic k-essence parameters $\alpha$, $A$ and $\nu$. Here, we impose a prior on the frequency $\nu$ of $\mathcal{U}[0, 3]$.}
    \label{fig:mke_lownu}
\end{figure}

Figure~\ref{fig:mke_lownu} show a triangle plot with k-essence parameter constraints for the low frequency prior. All of the frequency posteriors peak at $\nu \approx 1$. Interestingly, the $\alpha$ posterior is shifted towards negative values, indicating oscillations around $w < -1$. Finally, the posterior peaks towards high values of $A$. Except for the results using Pantheon, a zero amplitude value is excluded at above $2\sigma$ confidence level. These parameter combinations suggest a behavior similar to the preferred $w_0w_a$ behavior from the DESI results, in which the equation of state is below $-1$ in the past and starts oscillating slowly in the present, reaching values above $-1$. The constraints at $68\%$ confidence level are:
\begin{itemize}
    \item \makebox[4.1cm]{P18 + DESI + Pan:\hfill} $A > 0.405$, $\alpha = -0.37^{+0.32}_{-0.15}$, $\nu = 1.13\pm 0.62$;
    \item \makebox[4.1cm]{P18 + DESI + Pan+:\hfill} $A > 0.596$, $\alpha = -0.40^{+0.34}_{-0.23}$, $\nu = 1.11^{+0.52}_{-0.42}$;
    \item \makebox[4.1cm]{P18 + DESI + Uni3:\hfill} $A > 0.754$, $\alpha = -0.59^{+0.26}_{-0.36}$, $\nu = 1.28^{+0.41}_{-0.24}$;
    \item \makebox[4.1cm]{P18 + DESI + DESY5:\hfill} $A > 0.753$, $\alpha = -0.56^{+0.26}_{-0.36}$, $\nu = 1.23^{+0.36}_{-0.25}$.
\end{itemize}

\begin{figure}[h]
    \centering
    \includegraphics[width=0.95\textwidth]{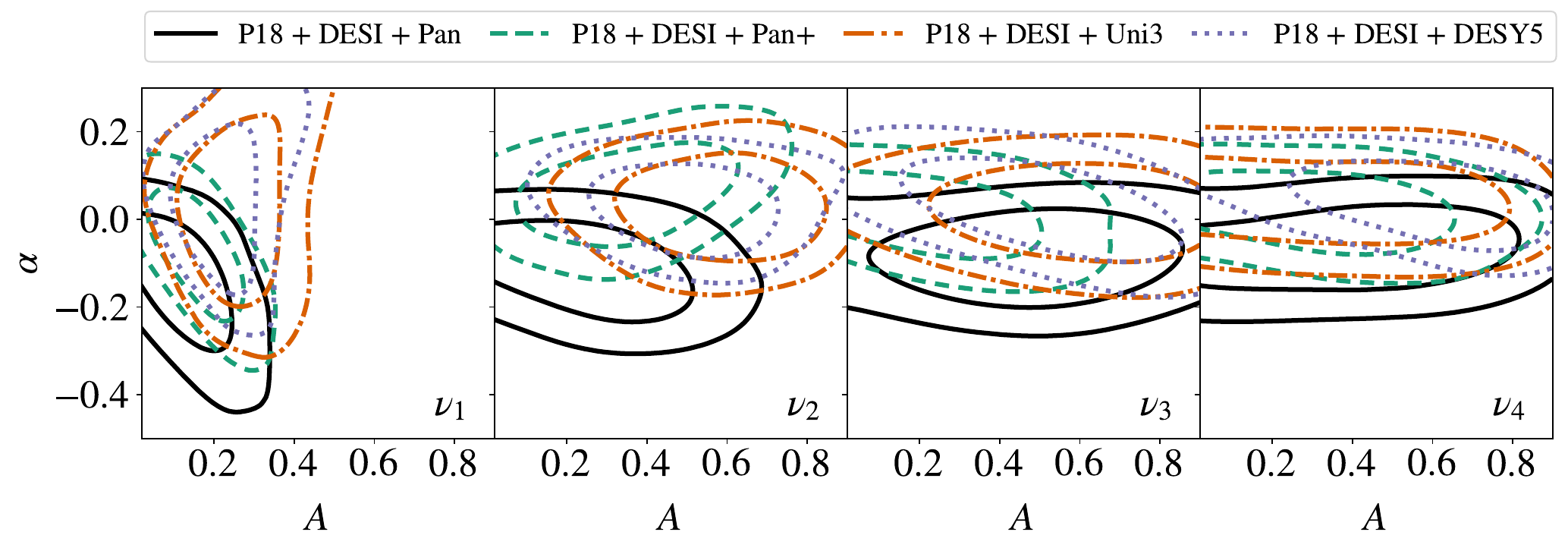}
    \caption{2D confidence contours (68\% and 95\%) of monodromic k-essence parameters $\alpha$ and $A$. Each panel represents a different $\nu$ prior (see Figure~\ref{fig:mke_freqs}) while each line type corresponds to a different dataset combination.}
    \label{fig:mke_allnu}
\end{figure}

Figure~\ref{fig:mke_allnu} shows 2D confidence contours for the MKE parameters $\alpha$ and $A$. For $\nu$ varying around $\nu_1$, all datasets present a mild preference for oscillations, $A > 0$, with constraints (68\% limits):
\begin{itemize}
    \item \makebox[4.1cm]{P18 + DESI + Pan:\hfill} $A = 0.146^{+0.060}_{-0.10}$;
    \item \makebox[4.1cm]{P18 + DESI + Pan+:\hfill} $A = 0.169^{+0.066}_{-0.088}$;
    \item \makebox[4.1cm]{P18 + DESI + Uni3:\hfill} $A = 0.266^{+0.081}_{-0.098}$;
    \item \makebox[4.1cm]{P18 + DESI + DESY5:\hfill} $A = 0.226^{+0.068}_{-0.084}$.
\end{itemize}
For $\nu$ varying around the next frequency $\nu_2$, the posteriors are shifted towards higher values of the amplitude $A$. Except for Pantheon, all dataset combinations exclude $A = 0$ with over $95\%$ confidence, with a strong preference for $A > 0$ when either Union3 or DESY5 SN are considered. The 1D marginalized constraints on $A$ are:
\begin{itemize}
    \item \makebox[4.1cm]{P18 + DESI + Pan:\hfill} $A = 0.29^{+0.14}_{-0.20}$;
    \item \makebox[4.1cm]{P18 + DESI + Pan+:\hfill} $A = 0.36^{+0.16}_{-0.19}$;
    \item \makebox[4.1cm]{P18 + DESI + Uni3:\hfill} $A = 0.58^{+0.20}_{-0.14}$;
    \item \makebox[4.1cm]{P18 + DESI + DESY5:\hfill} $A = 0.50^{+0.16}_{-0.15}$.
\end{itemize}

No significant preference for $A > 0$ is observed for the other frequency values, although the posterior distribution is compatible with high amplitude values. Since the energy density is an integral of $w(z)$, and the luminosity distance is an integral containing the energy density, high-frequency oscillations are averaged and $A$ has limited impact on the distance calculations. As for the $\alpha$ parameter, all datasets are consistent with $\alpha = 0$, suggesting oscillations around $w = -1$, see Eq. (\ref{eq:average_w_mke}).

\begin{figure}[!htbp]
    \centering
    \includegraphics[width=0.9\textwidth]{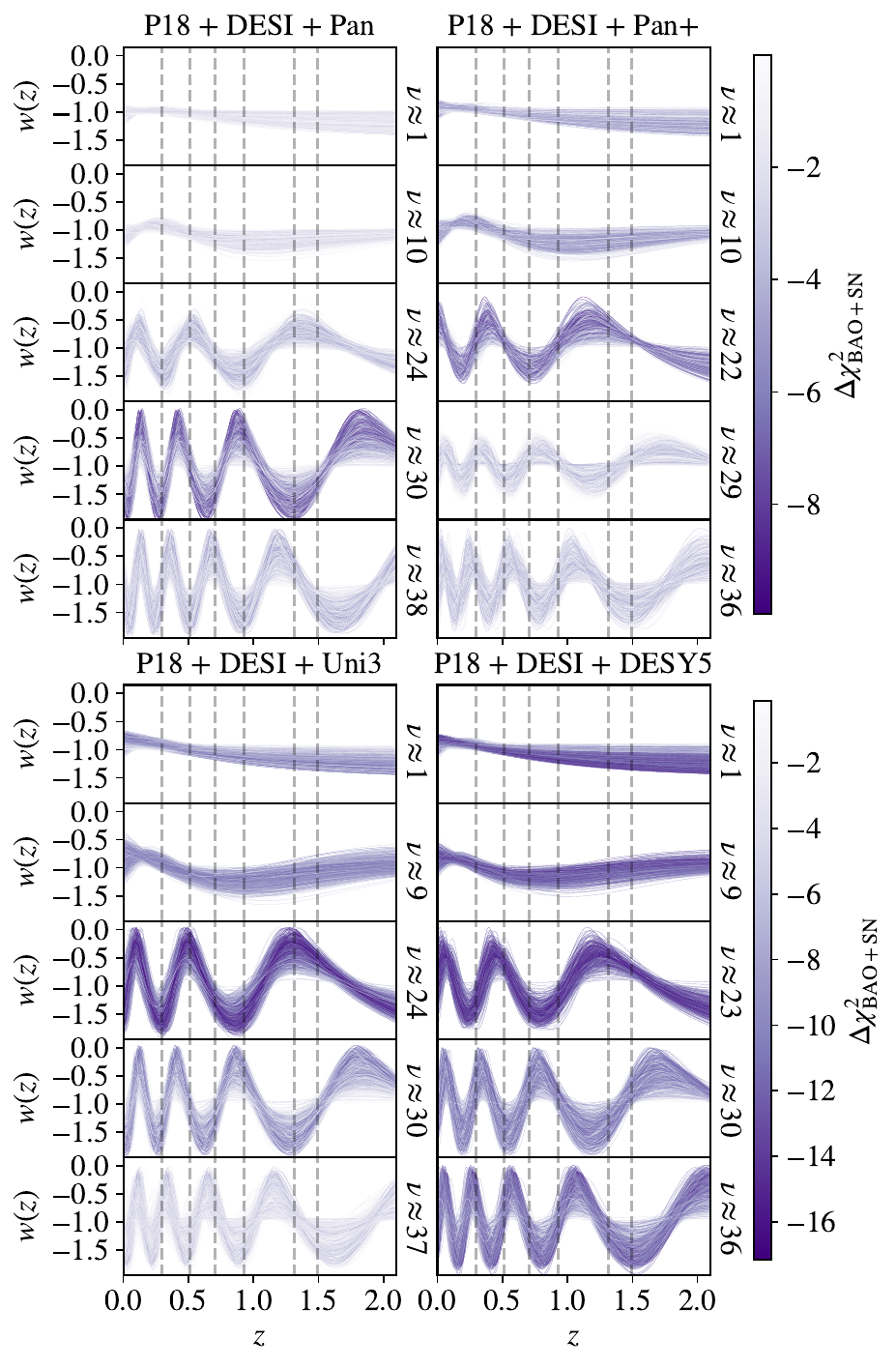}
    \caption{K-essence equation of state evolution for various MCMC samples of the model parameters $A$ and $\alpha$. Each panel corresponds to a different dataset combination and is divided into five subpanels, one for each frequency prior: $\nu \approx 1$ indicates $\nu \in [0, 3]$ and the other values indicate $\nu \in [\nu_i - 2, \nu_i + 2]$. Each $w(z)$ curve is color-coded by $\Delta \chi^2_\mathrm{BAO+SN} = \chi^2_\mathrm{BAO+SN}(\alpha, A, \nu) - \min\limits_{\Lambda \mathrm{CDM}}\chi^2_\mathrm{BAO+SN}$, with darker shades indicating lower values of $\Delta\chi^2_\mathrm{BAO+SN}$. The vertical dashed lines represent the BAO effective redshifts.}
    \label{fig:kessence_reconstr}
\end{figure}

To visualize the $w(z)$ behavior being preferred by the data, in Figure~\ref{fig:kessence_reconstr} we show $w(z)$ for different parameter samples in the MCMCs\footnote{We burn-in and thin the MCMC samples to ensure uncorrelated samples.}, color-coded by their $\chi^2_\mathrm{BAO+SN}$. The oscillation feature of $w(z)$ in this model is noticeable, and are more pronounced for the $\nu$ values that yield a lower $\chi^2$ according to Figure~\ref{fig:mke_freqs}. 

\subsection{Goodness-of-Fit}
\label{sec:chi2}
Using the MCMC samples, we can use a simulated annealing algorithm as described in e.g. \cite{procoli}, finding the global minimum of $\chi^2$ and the difference with the minimum $\chi^2$ obtained assuming $\Lambda$CDM. To assess which specific dataset drives the results, we report the $\chi^2$ value divided into three components: $(\chi^2_\mathrm{BAO}, \chi^2_\mathrm{SN}, \chi^2_\mathrm{CMB})$. For each dark energy model, we assess $\Delta \chi^2 = \chi^2_\mathrm{model} - \chi^2_{\Lambda\mathrm{CDM}}$, as a measure of the performance of the model to fit the data when compared to $\Lambda$CDM; the $\Delta\chi^2$ is divided into three components similarly. Table~\ref{tab:delta_chi2} shows the minimization results for each dark energy model and each dataset. For MKE, we perform minimizations starting at the minima for each MCMC with different $\nu$ prior and select the one that yields the minimum $\chi^2$.

\begin{table}[h]
\centering
\begin{tabular}{|l|l@{\hskip -.85in}l|l@{\hskip -1in}l|}
    \hline
        & {P18 + DESI + Pan} & & P18 + DESI + Pan+ &\\
    \hline 
        $\Delta \chi^2_{w_0w_a}$ & -2.5&(-0.7, 0.2, -2.0) & -7.0&(-2.1, -2.6, -2.2) \\
    \hline
        $\Delta \chi^2_\mathrm{trans}$ & -3.7&(-1.5, 0.3, -2.5)& -8.6&(-2.9, -3.3, -2.4) \\
    \hline
        \makecell{$\Delta \chi^2_{\mathrm{5-bin }\; w(z)}$} & -6.7&(-6.1, 1.0, -1.6) & -14.9&(-7.4, -5.8, -1.7) \\
    \hline
        \makecell{$\Delta \chi^2_{\mathrm{10-bin }\; w(z)}$} & -15.6&(-9.3, -5.2, -1.1) & -20.0&(-8.9, -9.7, -1.5) \\
    \hline
        $\Delta \chi^2_\mathrm{MKE}$ & -6.8&(-3.0, -4.4 +0.5) ($\nu \approx \nu_2$) & -7.2&(-2.6, -4.9, +0.4) ($\nu \approx \nu_2$) \\ 
    \hline
    \hline
        & P18 + DESI + Uni3 && P18 + DESI + DESY5 &\\
    \hline
        $\Delta \chi^2_{w_0w_a}$ & -11.9&(-3.0, -6.7, -2.2) & -15.4&(-4.0, -10.3, -1.0) \\ 
    \hline
        $\Delta \chi^2_\mathrm{trans}$ & -11.1&(-3.5, -6.2, -1.4) & -15.0&(-3.3, -11.3, -0.5) \\ 
    \hline
        \makecell{$\Delta \chi^2_{\mathrm{5-bin }\; w(z)}$} & -13.7&(-7.2, -5.2, -1.3) & -15.0&(-5.3, -8.6, -1.2) \\
    \hline
        \makecell{$\Delta \chi^2_{\mathrm{10-bin }\; w(z)}$} & -26.7&(-10.7, -14.3, -1.7) & -25.2&(-5.8, -20.3, +0.8)\\
    \hline
        $\Delta \chi^2_\mathrm{MKE}$ & -10.3&(-2.9, -6.7, -0.7) ($\nu \in [0, 3]$) & -13.8&(-4.1, -10.4, -0.7) ($\nu \in [0, 3]$) \\
    \hline
\end{tabular}
\caption{$\Delta\chi^2 = \chi^2_\mathrm{model} - \chi^2_{\Lambda\mathrm{CDM}}$ of the dark energy models, compared to the best-fit $\Lambda$CDM model. The value is broken down into $(\Delta\chi^2_\mathrm{BAO},\, \Delta\chi^2_\mathrm{SN},\, \Delta\chi^2_\mathrm{CMB})$. The reference $\chi^2_{\Lambda\text{CDM}}$ for each dataset combination is the format P18+DESI+``SNIa'' where ``SNIa'' refers to one of the Supernovae datasets from Pantheon:~$\chi^2 = 3804.6~(\chi^2_\mathrm{BAO}=13.6, \,\chi^2_\mathrm{SN}=1034.7 ,\chi^2_\mathrm{CMB}=2756.3)$; Pantheon+:~$\chi^2 = 4176.0~(\chi^2_\mathrm{BAO} = 14.4, \,\chi^2_\mathrm{SN}=1405.5, \,\chi^2_\mathrm{CMB} = 2756.1)$; Union3:~$\chi^2 = 2798.7~(\chi^2_\mathrm{BAO} =14.5, \,\chi^2_\mathrm{SN}=28.1, \,\chi^2_\mathrm{CMB} = 2756.1)$; and DESY5:~$\chi^2 =4418.7~(\chi^2_\mathrm{BAO}=15.9, \,\chi^2_\mathrm{SN}=1647.8, \,\chi^2_\mathrm{CMB} =2755.0)$.}
\label{tab:delta_chi2}
\end{table}

Despite having one additional parameter, the transition parametrization fits the BAO and SN datasets similarly to the $w_0w_a$ parametrization. For the analyses using Pantheon, Pantheon+ and Union3, the transition parametrization can fit DESI BAO slightly better than $w_0w_a$. We observe that the binned $w(z)$ parametrization can fit DESI BAO data considerably better than $w_0w_a$: using 5 bins, the binned $w(z)$ model can reduce $\chi^2_\mathrm{BAO}$ by $-5.4$, $-5.3$ and $-4.2$ when compared to $w_0w_a$ in the Pantheon, Pantheon+ and Union3 analyses, respectively. However, except for Pantheon+, the binned $w(z)$ with 5 bins cannot improve the fit to the SN datasets. For this reason, we conclude that the non-monotonic features observed in Figure~\ref{fig:binw_filtered} are driven by DESI BAO data rather than the SN datasets. Due to the increased generality, 10 redshift bins can fit both BAO and SN datasets better than all other models. Even in the P18 + DESI + DESY5 dataset, in which no other dark energy model could improve on the $w_0w_a$ fit, the 10-bins model showed an improvement of $\Delta \chi^2 = -9.8$ over $w_0w_a$. Finally, the MKE model performs similarly or better than $w_0w_a$ with respect to the SN fit: comparing $\chi^2_\mathrm{SN}$ between both models, MKE reduces $\chi^2_\mathrm{SN}$ by $-4.6$ considering Pantheon and $-2.3$ considering Pantheon+, with no significant difference in $\chi^2_\mathrm{SN}$ for the other SN datasets. However, the MKE model has a considerably worse fit to CMB than the other dynamical dark energy models: while most dynamical DE models present a $\Delta \chi^2_\mathrm{CMB}$ with respect to $\Lambda$CDM ranging from $-1$ to $-2.5$, MKE has a $\Delta\chi^2_\mathrm{CMB}$ ranging from $-0.7$ to $+0.5$.

To penalize models with an increased number of parameters, we transform the $\Delta \chi^2_\mathrm{model}$ value into a z-score, the same measure used by DESI to quote the $w_0w_a$ significance levels over $\Lambda$CDM \cite{desi_2024_bao}. Under the assumption that the likelihood is Gaussian, the quantity $\chi^2$ is a random variable following a $\chi^2$ distribution with $N_\mathrm{dv} - N_\mathrm{params}$ degrees of freedom, where $N_\mathrm{dv}$ is the number of elements in the data vector and $N_\mathrm{params}$ is the number of model parameters. Due to the additivity property of the of the $\chi^2$ distribution, the random variable $X = \chi^2_{\Lambda\mathrm{CDM}} - \chi^2_\mathrm{model}$ is also $\chi^2$-distributed with $\Delta N_\mathrm{params}$ degrees of freedom, where $\Delta N_\mathrm{params}$ is the difference in number of parameters between the dark energy model and $\Lambda$CDM. We have $\Delta N_\mathrm{params}^{w_0w_a} = 2$, $\Delta N_\mathrm{params}^{\mathrm{trans}} = 3$, $\Delta N_\mathrm{params}^{\mathrm{binw}} = 5 \text{ or } 10$, and $\Delta N_\mathrm{params}^{\mathrm{MKE}} = 3$. We then calculate $p$, the probability of $X > |\Delta \chi^2_\mathrm{model}|$. Defining $Z$ as another random variable following a standard normal distribution, the z-score is the value $z$ such that $P(-z < Z < z) = p$. We refrain, however, from interpreting the z-score as a statistical significance, only using this technique to quantitatively compare dark energy models while penalizing extra parameters. 

\begin{figure}[!htp]
    \centering
    \includegraphics[width=\textwidth]{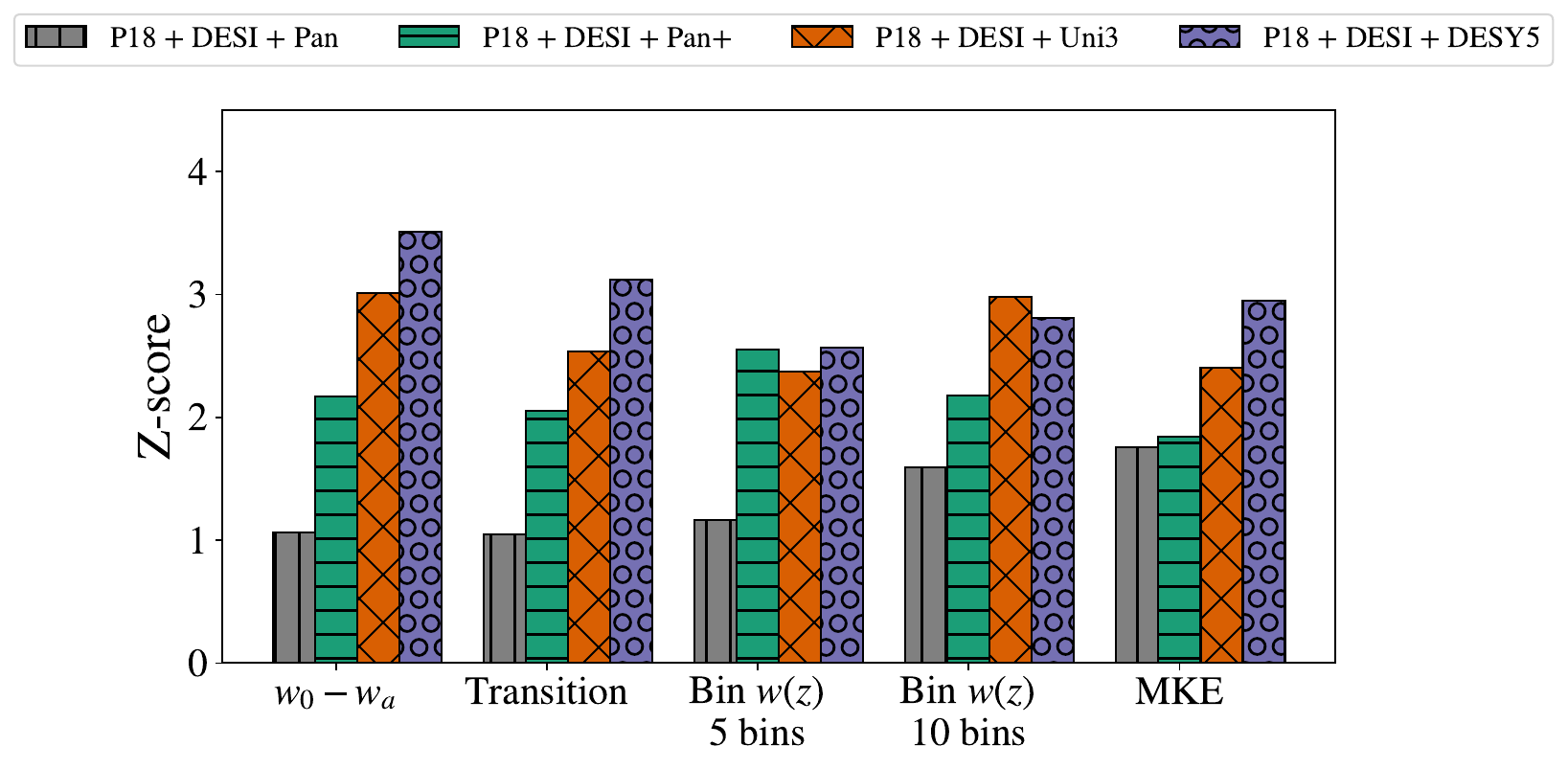}
    \caption{Z-score for each dark energy model and each dataset combination.}
    \label{fig:stats_sig}
\end{figure}

The z-score of each dark energy model is shown in Figure~\ref{fig:stats_sig}. We recover similar z-score for $w_0w_a$ reported by the DESI collaboration \cite{desi_2024_bao}. For Union3 and DESY5, $w_0w_a$ is the model with the highest z-score. Even though the binned $w(z)$ model improves the fit significantly, the increased amount of parameters overall amounts to a lower z-score. However, one could conceive a dark energy model that reproduces the filtered $w(z)$ behavior shown in Figure~\ref{fig:binw_filtered}: this model would have a similar $\Delta \chi^2$ to the binned $w(z)$ model but with less parameters, and therefore would be less penalized by statistical model comparison techniques. We also remark that the z-score of all models is below $2$ when Pantheon is considered, pointing to the fact that the discrepancy in $\Omega_m$ between DESI BAO and the newer SN datasets is a decisive factor for the observed dynamical dark energy preference~\cite{scalar_fields_desi}. 

Another way to penalize extra parameters is to use the Akaike Information Criterion, $\mathrm{AIC} = 2N_\mathrm{params} + \chi^2$. The preferred models are those with the minimum value of $\mathrm{AIC}$, representing a better fit with few parameters. For this purpose, we calculate $\Delta \mathrm{AIC}_\mathrm{model} = \mathrm{AIC}_\mathrm{model} - \mathrm{AIC}_{\Lambda\mathrm{CDM}}$ for each dataset and model, and the results are shown in Table~\ref{tab:aic}. For Union3 and DESY5, $w_0w_a$ is the model with the least $\mathrm{AIC}$, with $\Delta \mathrm{AIC} = -7.9$ and $-11.4$, respectively. For Pantheon+ the binned $w(z)$ with $N_\mathrm{bins} = 5$ has a lower $\mathrm{AIC}$, with $\Delta\mathrm{AIC} = -4.9$ compared to $\Delta\mathrm{AIC} = -3.0$ for $w_0w_a$. These results are consistent with those shown in Figure~\ref{fig:stats_sig}, indicating that $w_0w_a$ is still the most cost-effective model in terms of goodness-of-fit.

\begin{table}[h]
\centering
\begin{tabular}{|l|c|c|l|l|}
    \hline
        & P18 + DESI + Pan & P18 + DESI + Pan+ \\
    \hline
        $\Delta \mathrm{AIC}_{w_0w_a}$ & +1.5 & -3.0 \\
    \hline
        $\Delta \mathrm{AIC}_\mathrm{trans}$ & +2.3 & -2.6 \\
    \hline
        \makecell{$\Delta \mathrm{AIC}_{\mathrm{5-bin }\; w(z)}$} & +3.3
     & -4.9 \\
    \hline
        \makecell{$\Delta \mathrm{AIC}_{\mathrm{10-bin }\; w(z)}$} & +4.4 & 0.0 \\
    \hline
    $\Delta \mathrm{AIC}_\mathrm{MKE}$ & -0.8 ($\nu \approx \nu_2$) & -1.2 ($\nu \approx \nu_2$) \\ 
    \hline
    \hline
        & P18 + DESI + Uni3 & P18 + DESI + DESY5 \\
    \hline
        $\Delta \mathrm{AIC}_{w_0w_a}$ & -7.9 & -11.4 \\ 
    \hline
        $\Delta \mathrm{AIC}_\mathrm{trans}$ & -5.1 & -9.0 \\ 
    \hline
        \makecell{$\Delta \mathrm{AIC}_{\mathrm{5-bin }\; w(z)}$} & -3.7& -5.0 \\
    \hline
        \makecell{$\Delta \mathrm{AIC}_{\mathrm{10-bin }\; w(z)}$} & -6.7& -5.2 \\
    \hline
        $\Delta \mathrm{AIC}_\mathrm{MKE}$ & -4.3 ($\nu \in [0, 3]$) & -7.8 ($\nu \in [0, 3]$)\\
    \hline
\end{tabular}
\caption{$\Delta\mathrm{AIC} = \Delta\chi^2_\mathrm{model} + 2\Delta N_\mathrm{params}$ of the dark energy models, compared to the best-fit $\Lambda$CDM model.}
\label{tab:aic}
\end{table}

\subsection{Constraints on \texorpdfstring{$\sum m_\nu$}{}}
\label{sec:mnu}

Hitherto, our analysis has assumed two massless neutrinos and one massive neutrino with $\sum m_\nu = 0.06$ eV, i.e.~the minimal value for the sum mass in the normal hierarchy scenario. This assumption is recurrent in the literature \cite{planck_2018_cmb,act_dr6_lensing_cp,desi_2024_bao} when $\sum m_\nu$ is kept fixed. In this Section, we consider the neutrino mass sum $\sum m_\nu$ as a free parameter, assuming three massive neutrinos and a degenerate hierarchy with priors shown in Table \ref{tab:priors}. For these constraints, it is interesting to assess the difference among the CMB datasets and the effect of including CMB lensing data. The constraints on $\sum m_\nu$ are assessed for different dark energy equations of state models of $w=-1$, $w_0w_a$, and binned $w(z)$ models.

\begin{table}[h]
\centering
\begin{tabular}{|l|l|l|l|l|l|}
\hline
\multicolumn{5}{|c|}{Upper limit ($95\%$ C. L.) on $\sum m_\nu$ (eV)} \\
\hline
\hfill P18 + DESI + & Pan & Pan+ & Uni3 & DESY5 \\
\hline
$\Lambda\mathrm{CDM}$ & 0.10 & 0.11 & 0.11 &  0.12 \\
\hline
${w_0w_a}$ & 0.18 & 0.17  & 0.21 &  0.19\\ 
\hline
{${\mathrm{5-bin }\; w(z)}$} & 0.19 & 0.19  & 0.19 &  0.18\\ 
\hline
{${\mathrm{10-bin }\; w(z)}$}& 0.16 & 0.16  & 0.16 &  0.15\\
\hline
\hline
\hfill ACT+P650 + DESI + & Pan & Pan+ & Uni3 & DESY5\\
\hline
${\Lambda\mathrm{CDM}}$ & {0.16} &{0.18} & {0.18} & {0.20}\\
\hline
${w_0w_a}$ & {0.31} & {0.28} & {0.34} & {0.32}\\
\hline
{${\mathrm{5-bin }\; w(z)}$} & {0.32} &  {0.33} & {0.32} & {0.31}\\
\hline
{${\mathrm{10-bin }\; w(z)}$} & {0.26} & {0.26} & {0.26} & {0.24}\\
\hline
\hline
\hfill ACT+P650 + ACTL + DESI + & Pan & Pan+ & Uni3 & DESY5\\
\hline
${\Lambda\mathrm{CDM}}$ & {0.10}  & {0.11} & {0.11} &  {0.12}\\
\hline
${w_0w_a}$ & {0.22}& {0.21} & {0.24} & {0.23}\\
\hline
{${\mathrm{5-bin }\; w(z)}$} & {0.23} & {0.23} & {0.23} &  {0.22}\\
\hline
{${\mathrm{10-bin }\; w(z)}$} & {0.17} &  {0.18} & {0.17} &  {0.17}\\
\hline

\end{tabular}
\caption{Upper limit on the sum of the neutrino mass at 2$\sigma$ for different dataset combinations and dark energy models.}
\label{tab:mnu}
\end{table}

\begin{figure}[!htbp]
    \centering
    \includegraphics[width=0.95\linewidth]{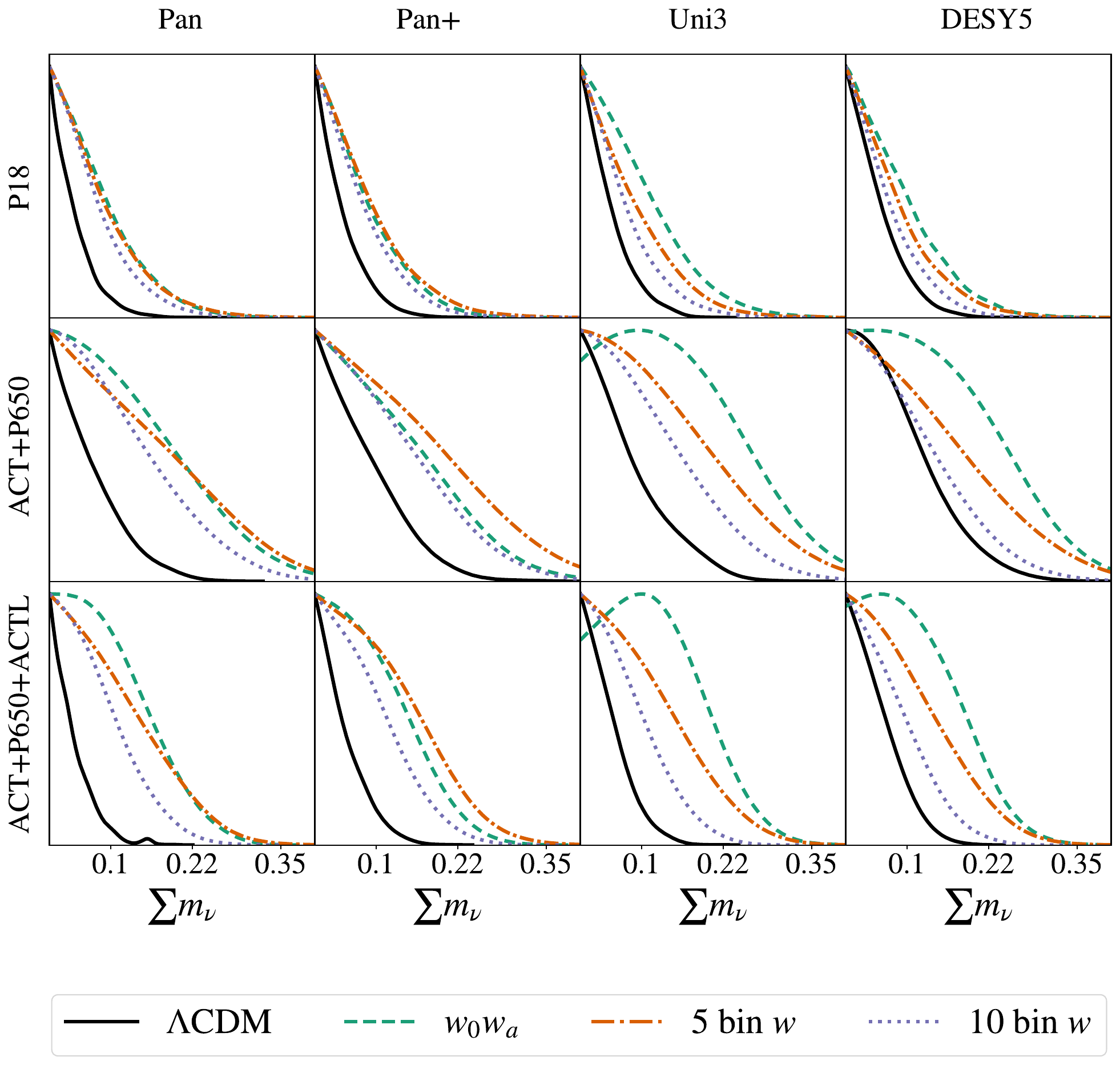}
    \caption{1D marginalized posteriors for the neutrino mass sum $\sum m_\nu$ for datasets and models presented in Table \ref{tab:mnu}. Each panel represents a different dataset combination with BAO from DESI. The top row represents constraints with Planck 2018, the middle row shows constraints using ACT DR4 + Planck 2018 TT TE EE ($\ell < 650$), and the bottom row is the same as the middle row with the addition of CMB lensing from ACT DR6. Each column has a different supernovae catalog: Pantheon, Pantheon+, Union3, and DESY5.}
    \label{fig:mnu}
\end{figure}

The results are illustrated in Figure~\ref{fig:mnu}, which shows the marginalized constraints on the neutrino mass sum assuming different dark energy models; Table~\ref{tab:mnu} shows the $95\%$ upper limit. Previous works have shown that constraints on $\sum m_\nu$ under the $\Lambda$CDM model prefer negative neutrino masses \cite{negative_neutrino_mass}. This effect is aggravated by the ``lensing anomaly" present on the Planck 2018 temperature and polarization power spectra in the \texttt{Plik} likelihood\footnote{The Planck collaboration has released two likelihoods in which the lensing anomaly is diminished: \texttt{CamSpec} and \texttt{HiLLiPop}. Analyses using the new likelihoods show relaxed constraints on $\sum m_\nu$ when compared to the original \texttt{plik} likelihood \cite{desi_planck_pr4, mnu_edge}, but the preference for negative neutrino masses persist.}: the lensing of the spectra exceeds the theoretical $\Lambda$CDM lensing effect \cite{planck_2018_cmb}. Extra lensing would be caused by an increased formation of structure. Therefore, the anomaly makes Planck 2018 data disfavor mechanisms that suppress the growth of structure; in particular, this anomaly causes a preference for even more negative neutrino masses, and the Bayesian posterior upper limit, assuming $\sum m_\nu > 0$, gets lower. ACT CMB data does not present a lensing anomaly \cite{act_cmb}, and we observe an increase in the upper bound on $\sum m_\nu$ compared to Planck 2018 \cite{desi_planck_pr4}. The constraints obtained with the ACT+P650 dataset are wider by a factor of $1.5-1.9$ compared to P18. However, including ACT DR6 lensing data puts the constraints on the same level as P18.

Assuming the $w_0w_a$ parametrization, the $\sum m_\nu$ constraints are relaxed by a factor of 1.8-2.2 compared to $\Lambda$CDM,  depending on the dataset considered. The binned $w(z)$ with 5 redshift bins shows similar constraints to $w_0w_a$; however, although having an increased number of parameters, the case with 10 redshift bins has slightly tighter constraints on $\sum m_\nu$ than $w_0w_a$ by $75\%-90\%$. One might expect that a more general model with many more parameters would yield relaxed parameter constraints; however, if the binned $w(z)$ model makes the preferred value of $\sum m_\nu$ become more negative than that of $w_0w_a$, the Bayesian posterior gets tighter as a consequence, since the physical prior excludes negative neutrino masses\footnote{We thank Daniel Green for this remark.}. Our current results cannot confirm such hypothesis, and we leave this for future investigations. A similar effect has been noted in \cite{mnu_constraints_tight}, where the $w_0w_a$ dark energy parametrization with priors such that $w_0 > -1$ and $w_0 + w_a > -1$ has tighter constraints on $\sum m_\nu$ than $\Lambda$CDM.

\section{Conclusion}
\label{sec:conclusion}
DESI 2024 BAO data, combined with recent supernovae data from Pantheon+, Union3, and DESY5 catalogs, has shown an intriguing hint in favor of a dynamical dark energy equation of state. Analysis of the $w_0w_a$ parametrization, where $w(z)$ is a linear function of the scale factor, favors a transition from phantom dark energy at redshifts $z > 0.5$ to a quintessence behavior with $w > -1$ for $z < 0.5$. This exciting result, if confirmed, puts even more pressure on the $\Lambda$CDM model, and several studies have investigated this result using more general models that go beyond the simple $w_0w_a$ parametrization. 
Another surprising result from DESI data is the constraints on the neutrino mass sum, whose upper limit in the $\Lambda$CDM model is not far from the minimum possible value allowed by neutrino oscillation experiments, $\sum m_\nu > 0.06~\mathrm{eV}$. If the trend persists, the next generation of large-scale structure probes, such as Euclid and LSST, could put $\Lambda$CDM in tension with laboratory experiments. As cosmological data analyses show correlations between the inferred dark energy equation of state and neutrino mass sum, those constraints should be reassessed considering more general dark energy models.

To investigate the dynamical dark energy behavior preferred by DESI data, we have analyzed three general dark energy models: a transition parametrization, a model-agnostic parametrization with a binned $w(z)$, and an oscillating k-essence model. We investigate the dark energy parameter constraints from each dataset, assessing the preferred forms of $w(z)$ and pointing out differences and similarities among each supernovae catalog, especially between Pantheon and the other catalogs. We found that, for the transition parametrization, the transition redshift $z_c$ is weakly constrained, and the overall $w(z)$ behavior is compatible with those obtained from a $w_0w_a$ analysis. For the binned $w(z)$, we tested two choices of redshift binning, one with 5 bins and the other with 10 bins, and we performed a principal component analysis of the obtained MCMC samples. While the error bars on the equation of state $w(z_i)$ are large due to the increased number of parameters, we found a preference for non-monotonic principal components in both choices of redshift bins, driven by DESI data. To assess the non-trivial behavior preferred by data, we have filtered the principal components with nonzero amplitude means, finding an interesting non-monotonic signal.
Furthermore, the thawing behavior discussed in reference~\cite{scalar_fields_desi} is parametrized by a single principal component, and all SN datasets prefer it except for Pantheon. For the monodromic k-essence model, we have minimized $\chi^2$ for fixed values of frequency $\nu$, finding the values that yield a better fit to the data. Among the different SN datasets, the preferred frequency values are very similar, all showing evidence for a nonzero amplitude.

To assess the goodness-of-fit of each model, we have minimized the $\chi^2$ for each dark energy model analyzed in the MCMC explorations, using the simulated annealing technique. Overall, the transition parametrization fits all datsets we investigated as well as $w_0w_a$, despite having one extra parameter. The binned $w(z)$ parametrization can improve the fit to DESI BAO significantly. The monodromic k-essence model, while having competitive goodness-of-fit to BAO and SN datasets, cannot improve the fit to CMB data. We penalize the increased parameters by translating the $\Delta \chi^2$ into a z-score and calculating the Akaike Information Criterion. Among all dark energy models, $w_0w_a$ is the most cost-effective model, significantly improving the fit of all datasets with only two additional parameters compared to $\Lambda$CDM. However, a theoretical dark energy model that reproduces the dynamics shown in Figure~\ref{fig:binw_filtered} with fewer parameters than the binned $w(z)$ would be less penalized by model comparison techniques.

Furthermore, we have investigated the constraints on the neutrino mass sum $\sum m_\nu$ considering four different dark energy models: the cosmological constant $\Lambda$, $w_0w_a$, and the model-agnostic binned $w(z)$ with 5 and 10 redshift bins. The tight constraints on $\sum m_\nu$ under the $\Lambda$CDM model are greatly relaxed when $w_0w_a$ or the binned $w(z)$ with 5 bins are assumed, while the two latter models show similar constraints. For the case with 10 bins, we observe an overall tightening of the constraints by $70\%-90\%$ depending on the dataset combination.

Future DESI releases \cite{desi_galaxies_qso}\footnote{During the review of this article, DESI has released its cosmological constraints from the full-shape power spectrum~\cite{desi_full_shape}, which agrees with the BAO-only analysis.}, as well as BAO distance measurements from Euclid \cite{euclid_cosmology,euclid_mission_overview}, galaxy clustering, weak lensing and cross-correlation, as forecasted by Euclid \cite{euclid_prep_forecast_val,euclid_prep_forecast_cosmo}, in combination with future supernovae observations, such as from the Roman Space Telescope \cite{roman_sn, roman_multiprobe} and the Vera C. Rubin Observatory \cite{lsst_srd} will be able to pinpoint the expansion history and the behavior of dark energy at low redshifts with increased precision. Determining the low redshift dark energy dynamics with general parametrizations is extremely important in the current era of precision cosmology: it can shed light on the standing cosmological tensions and point towards fundamental dark energy models that can reproduce the observed features.

\acknowledgments

While writing this article, the authors became deeply saddened by the passing of Prof. Eric Baxter. May his memory be a blessing.
We thank Dillon Brout for valuable discussions about the supernova datasets. We also thank Daniel Green for discussions regarding neutrino mass constraints. The authors would also like to thank the Stony Brook Research Computing and Cyberinfrastructure, and the Institute for Advanced Computational Science at Stony Brook University for access to the high-performance SeaWulf computing system, which was made possible by a \$1.4M National Science Foundation grant (\#1531492). The authors further acknowledge that this research was supported by high-performance computing resources supplied by the Center for Scientific Computing (NCC/GridUNESP) of the São Paulo State University (UNESP). DHFS acknowledges the financial support from Coordination for the Improvement of Higher Education Personnel (CAPES) and the Institutional Internationalization Program (PrInt), JR acknowledges the financial support from São Paulo Research Foundation (FAPESP), under grant \#2022/13999-5. The work of RR is supported in part by the São Paulo Research Foundation (FAPESP) through grant \# 2021/10290-2 and by the Brazilian Agency CNPq through a productivity grant \# 311627/2021-8 and the INCT of the e-Universe.

\appendix
\section{Full Reconstruction of the Equation of State}
\label{app:binw_full}
\begin{figure}[!htbp]
    \centering
    \includegraphics[width=\textwidth]{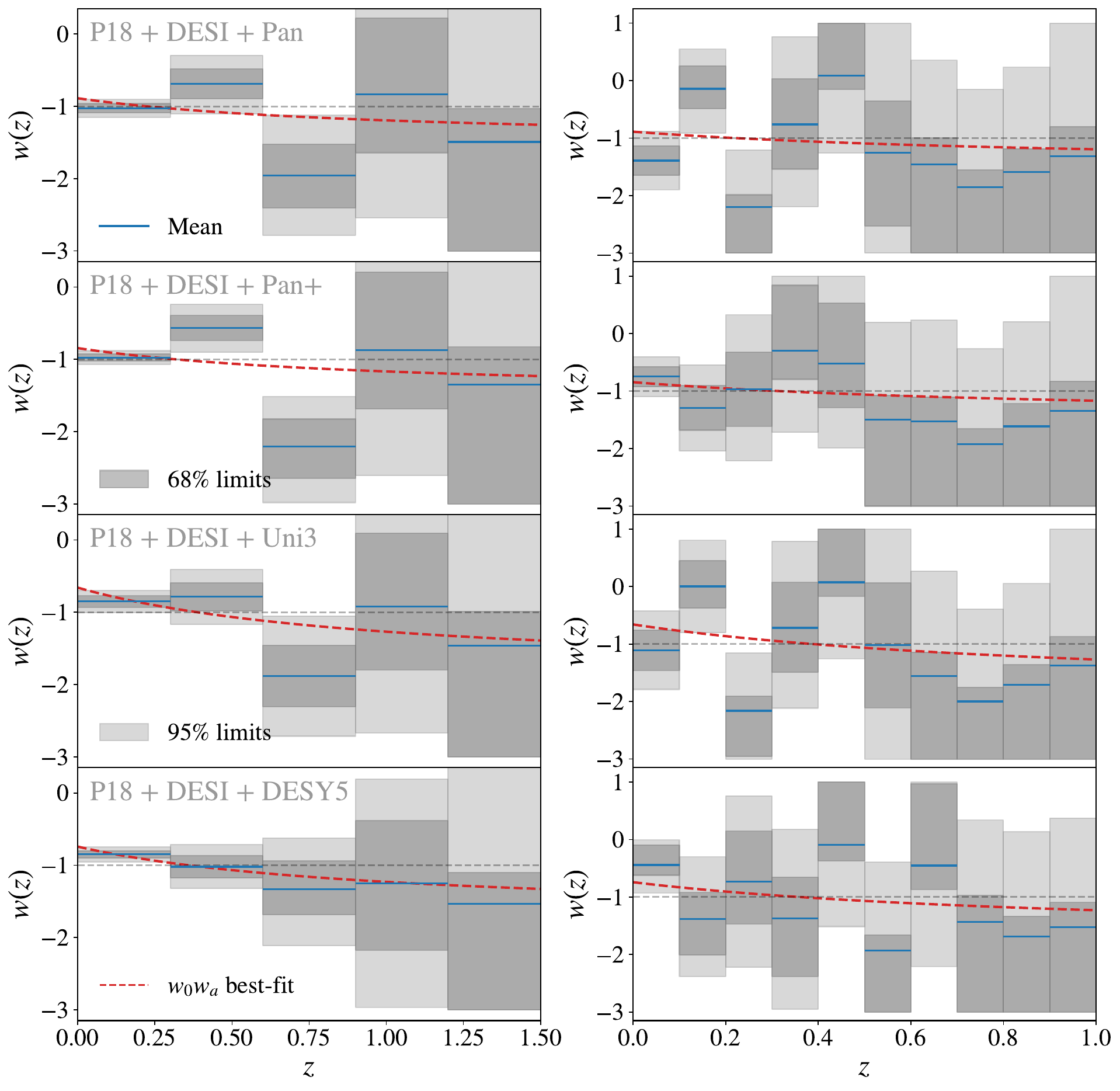}
    \caption{Reconstruction of the dark energy equation of state using the binned $w(z)$, Equation~\ref{eq:binw}, for the different SN datasets. The blue line indicates the mean value, and the darker and lighter gray bands indicate $68\%$ and $95\%$ limits. We also include the best-fit $w_0w_a$ model as a red dashed line for comparison purposes. The left panels show the reconstruction for $N_\mathrm{bins} = 5$, while the right panels show the filtered constraints for $N_\mathrm{bins} = 10$. Each row corresponds to a dataset combination including a different SN catalog.}
    \label{fig:binw_reconstr}
\end{figure}

In Figure~\ref{fig:binw_filtered}, we have presented the filtered constraints on $w_i$, the model-agnostic dark energy parameters in Equation~\ref{eq:binw}. In this appendix, for completeness, we show the unfiltered reconstruction, i.e., the full reconstruction for $w(z)$ in both cases of 5 and 10 bins inferred directly from the $w_i$ samples. The result is shown in Figure~\ref{fig:binw_reconstr}.

We also classify the binned $w(z)$ MCMC samples into different categories according to the bins in which the first phantom crossing occurred and in Tables~\ref{tab:crossing_percentages_5} and \ref{tab:crossing_percentages_10} we show the percentages of MCMC samples in each class.

\begin{table}[h]
\centering
\begin{tabular}{|c|c|c|c|c|}
    \hline
        & \multicolumn{4}{c|}{P18 + DESI +} \\
    \hline
        & Pan & Pan+ & Uni3 & DESY5 \\
    \hline 
        $1 \rightarrow 2$ & $70.1\%$ & $30.8\%$ & $17.0\%$ & $55.5\%$ \\
    \hline
        $2 \rightarrow 3$ & $28.5\%$ & $69.1\%$ & $82.6\%$ & $41.9\%$ \\
    \hline
        $3 \rightarrow 4$ & $0.82\%$ & $0.06\%$ & $0.5\%$ & $2.4\%$ \\
    \hline
        $4 \rightarrow 5$ & $0.04\%$ & $0.03\%$ & $0.03\%$ & $0.2\%$ \\
    \hline
        No crossing & $0.02\%$ & $0$ & $0$ & $0$ \\
    \hline
\end{tabular}
\caption{Percentages of 5-bin $w(z)$ MCMC samples that present the first phantom crossing between the bins. After burn-in, each MCMC contains around 300000 samples.}
\label{tab:crossing_percentages_5}
\end{table}

\begin{table}[h]
\centering
\begin{tabular}{|c|c|c|c|c|}
    \hline
        & \multicolumn{4}{c|}{P18 + DESI +} \\
    \hline
        & Pan & Pan+ & Uni3 & DESY5 \\
    \hline 
        $1 \rightarrow 2$ & $94.1\%$ & $83.4\%$ & $63.3\%$ & $79.5\%$ \\
    \hline
        $2 \rightarrow 3$ & $5.7\%$ & $15.7\%$ & $35.1\%$ & $17.3\%$ \\
    \hline
        $3 \rightarrow 4$ & $0.2\%$ & $0.6\%$ & $1.5\%$ & $3.0\%$ \\
    \hline
        $4 \rightarrow 5$ & $0$ & $0.2\%$ & $0.02\%$ & $0.1\%$ \\
    \hline
        $5 \rightarrow 6$ & $0$ & $0.1\%$ & $0.04\%$ & $0.1\%$ \\
    \hline
        $6 \rightarrow 7$ & $0$ & $0$ & $0.02\%$ & $0$ \\
    \hline
        $7 \rightarrow 8$ & $0$ & $0$ & $0$ & $0$ \\
    \hline
        $8 \rightarrow 9$ & $0$ & $0$ & $0$ & $0$ \\
    \hline
        $9 \rightarrow 10$ & $0$ & $0$ & $0$ & $0$ \\
    \hline
        No crossing & $0$ & $0$ & $0$ & $0$ \\
    \hline
\end{tabular}
\caption{Percentages of 10-bin $w(z)$ MCMC samples that present the first phantom crossing between the bins. After burn-in, each MCMC contains around 500000 samples.}
\label{tab:crossing_percentages_10}
\end{table}

\section{Continuous Binned \texorpdfstring{$w(z)$}{}}\label{app:smooth_binw}
The binned $w(z)$ model presented in Section~\ref{sec:binw} has discontinuities in the equation of state. To ensure that such discontinuities do not invalidate the model, we implement a continuous version, in which the equation of state is parametrized as
\begin{equation}\label{eq:smooth_binw}
    w(z) = \sum_{i=1}^{N}\left\{w_{i-1} + \frac{w_{i} - w_{i-1}}{2}\left[ 1 + \tanh\left(\frac{z - z_i}{\sigma} \right) \right] \right\}.
\end{equation}
This parametrization preserves the same parameters $z_i$ and $w_i$ of the discontinuous case, while adding one extra parameter $\sigma$ controlling the width of this transition. The smaller the widths are the closer this model is to the original binned $w(z)$ model. We have post-processed the binned $w(z)$ MCMC samples with this new model, assessing the difference in $\chi^2$ between both cases. We have chosen to fix the $\sigma$ parameter to three values of $\sigma = 0.01$, $\sigma = 0.03$ and $\sigma = 0.05$.

\begin{figure}
    \centering
    \includegraphics[width=0.7\linewidth]{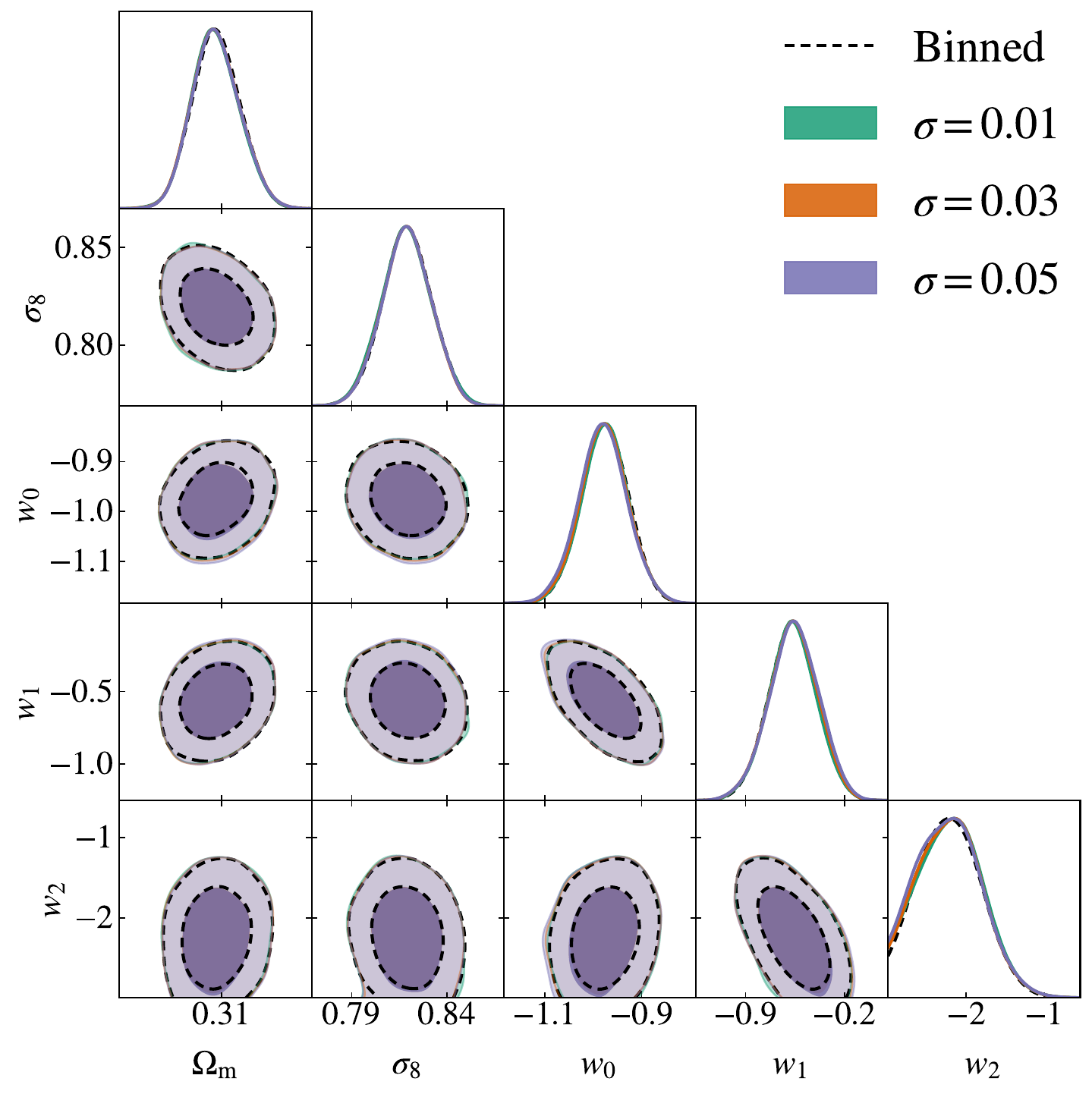}
    \caption{2D parameter constraints for the binned $w(z)$ model and the continuous version with three values of $\sigma = 0.01$, $0.03$ and $0.05$. We observe no significant differences between the constraints.}
    \label{fig:binw_comparison_mcmc}
\end{figure}

Figure~\ref{fig:binw_comparison_mcmc} shows the constraints on $\Omega_m$, $\sigma_8$, $w_0$, $w_1$ and $w_2$ for the discontinuous case and the three continuous cases. We observe almost perfect agreement between all cases.

\begin{figure}[!htbp]
    \centering
    \includegraphics[width=0.5\linewidth]{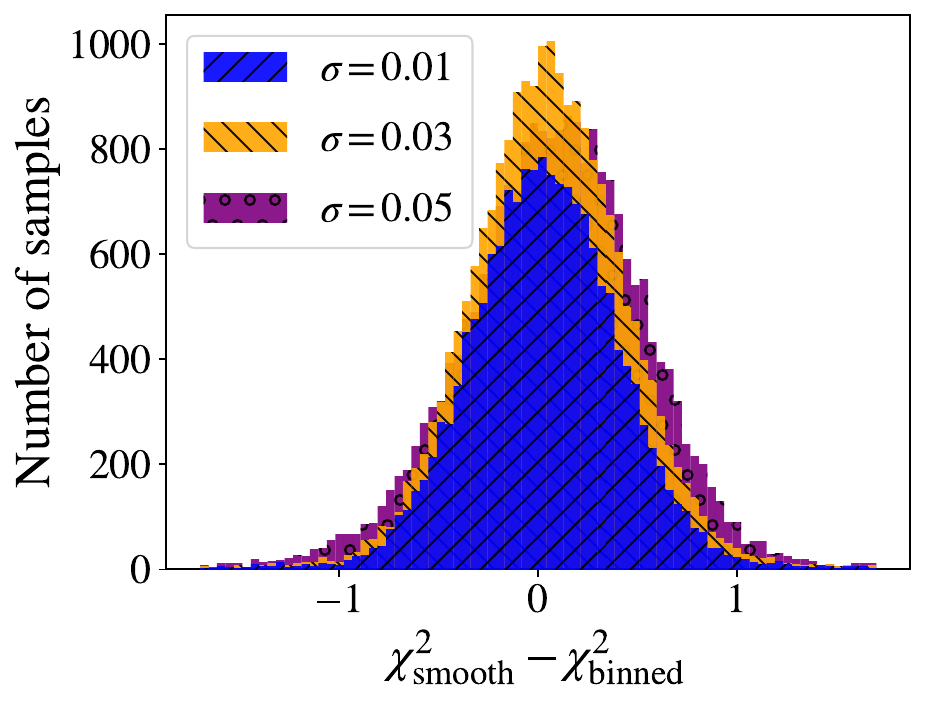}
    \caption{Histogram of $\chi^2$ differences between the binned $w(z)$ model described in Section~\ref{sec:binw} and the continuous version in Equation~\ref{eq:smooth_binw}}
    \label{fig:smooth_binw_hist}
\end{figure}

A histogram of $\chi^2$ differences is shown in Figure~\ref{fig:smooth_binw_hist} for the three $\sigma$ values. For $\sigma = 0.01$ and $\sigma = 0.03$, the percentage of points with $|\Delta \chi^2| < 1$ is $96\%$, falling to $94\%$ for $\sigma = 0.05$. The percentages of points within a threshold of $|\Delta\chi^2| < 0.5$ are $81\%$, $79\%$ and $71\%$ for the respective $\sigma$ values.

\section{Transition Model: Varying \texorpdfstring{$\sigma$}{}}\label{app:tanh_varysigma}
\begin{figure}[!htbp]
    \centering
    \includegraphics[width=0.5\linewidth]{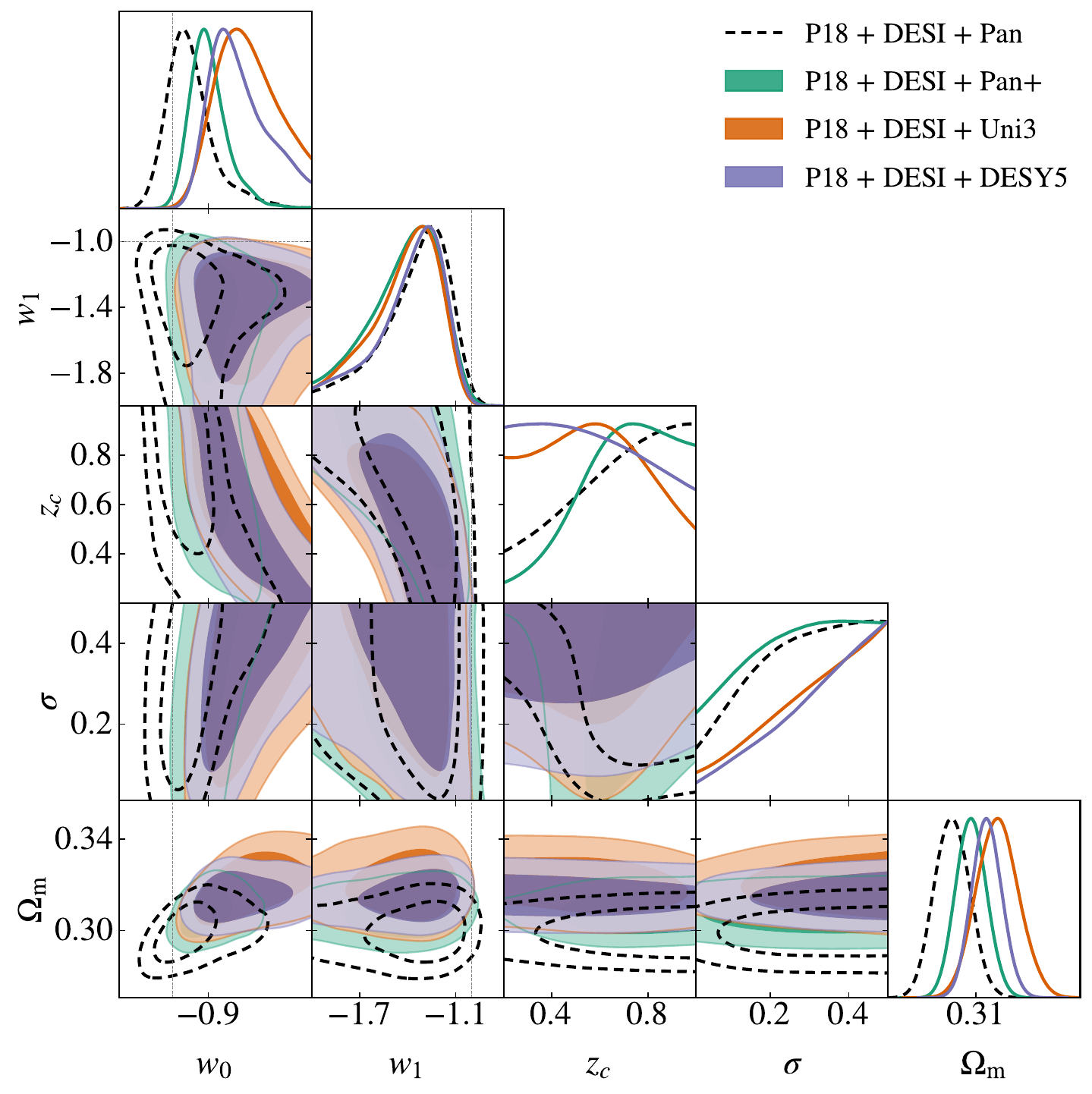}
    \caption{Parameter constraints for the transition model described in Section~\ref{sec:tanh}. Here, we vary the $\sigma$ parameter with a prior $\mathcal{U}[0.01, 0.5]$.}
    \label{fig:tanh_varysigma}
\end{figure}

In this Appendix, we show parameter constraints for the transition model described in Section~\ref{sec:tanh}, but now varying the $\sigma$ parameter with a prior $\mathcal{U}[0.01, 0.5]$. The results are shown in Figure~\ref{fig:tanh_varysigma}.

\newpage
\bibliographystyle{JHEP}
\bibliography{paper.bib}

\end{document}